\documentclass[reqno]{amsart}
\usepackage{epsf}
\usepackage{color}
\def\be{\begin{equation}}
\def\ee{\end{equation}}
\def\ba{\begin{array}}
\def\ea{\end{array}}
\def\bea{\begin{eqnarray}}
\def\eea{\end{eqnarray}}

\def\erm{{\mathrm e}}

\def\rt{\rightarrow}

\def\bvec{\mathbf}
\def\ov{\overline}
\def\prt{\partial}
\newcommand{\grbf}{\mbox{\boldmath $\nabla$}}

\def\gr{\grbf \,}

\def\gr{\mbox{\boldmath $\nabla$} }

\def\pbvec{\mbox{\boldmath $p$}}

\def\xbvec{\mbox{\boldmath $x$}}
\def\alphavec{\mbox{\boldmath $\alpha$}}
\def\sigmavec{\mbox{\boldmath $\sigma$}}
\def\psivec{\mbox{\boldmath $\psi$}}

\newcommand{\nc}{\newcommand}
\nc{\tcb}{\textcolor{blue}}
\nc{\tcr}{\textcolor{red}}
\nc{\tcg}{\textcolor{green}}

\nc{\qq}{\qquad\qquad}
\nc{\dis}{\displaystyle}
\nc{\ug}{\; = \;}
\nc{\Ebf}{\mbox{\boldmath $E$}}
\nc{\Bbf}{\mbox{\boldmath $B$}}
\nc{\abf}{\mbox{\boldmath $a$}}
\nc{\vbf}{\mbox{\boldmath $v$}}
\nc{\Fbf}{\mbox{\boldmath $F$}}
\nc{\rbf}{\mbox{\boldmath $r$}}
\nc{\Jbf}{\mbox{\boldmath $J$}}
\nc{\rd}{{\rm d}}
\nc{\dtau}{{\rd\tau}}
\nc{\dt}{{\rd t}}
\nc{\omf}{{\frac{\omega}{\omz}}}
\nc{\tz}{\tau_0}
\newcommand{\alphabf}{\mbox{\boldmath $\alpha$}}
\newcommand{\sigmabf}{\mbox{\boldmath $\sigma$}}

\begin{document}


\vspace{1truecm}

\title[Searching for an equation: Dirac, Majorana and the {\it others}]{Searching for an equation: \\ Dirac, Majorana and the {\it others}}

\author{S. Esposito}
\address{{\it S. Esposito}: I.N.F.N. Sezione di
Napoli, Complesso Universitario di M. S. Angelo, Via Cinthia,
80126 Naples, Italy ({\rm Salvatore.Esposito@na.infn.it})}%

\begin{abstract}
We review the non-trivial issue of the relativistic description of a quantum mechanical system that, contrary to a common belief, kept theoreticians busy from the end of 1920s to (at least) mid 1940s. Starting by the well-known works by Klein-Gordon and Dirac, we then give an account of the main results achieved by a variety of different authors, ranging from de Broglie to Proca, Majorana, Fierz-Pauli, Kemmer, Rarita-Schwinger and many others. 

A particular interest comes out for the general problem of the description of particles with {\it arbitrary} spin, introduced (and solved) by Majorana as early as 1932, and later reconsidered, within a different approach, by Dirac in 1936 and by Fierz-Pauli in 1939. The final settlement of the problem in 1945 by Bhabha, who came back to the general ideas introduced by Majorana in 1932, is discussed as well, and, by making recourse also to unpublished documents by Majorana, we are able to reconstruct the line of reasoning behind the Majorana and the Bhabha equations, as well as its evolution. Intriguingly enough, such an evolution was {\it identical} in the two authors, the difference being just the period of time required for that: probably few weeks in one case (Majorana), while more than ten years in the other one (Bhabha), with the contribution of several intermediate authors.

The important unpublished contributions by Majorana anticipated later results obtained, in a more involved way, by de Broglie (1934) and by Duffin and Kemmer (1938-9), and testify the intermediate steps in the line of reasoning that led to the paper published in 1932 by Majorana, while Bhabha took benefit of the corresponding (later) published literature. Majorana's paper of 1932, in fact, contrary to the more complicated Dirac-Fierz-Pauli formalism, resulted to be very difficult to fully understand (probably for its pregnant meaning and latent physical and mathematical content): as is clear from his letters, even Pauli (who suggested its reading to Bhabha) took about one year in 1940-1 to understand it. This just testifies for the difficulty of the problem, and for the depth of Majorana's reasoning and results.

The relevance for present day research of the issue here reviewed is outlined as well.
\end{abstract}

\maketitle



\section{Introduction}

\noindent The birth of quantum mechanics was driven by the basic principles of the theory of relativity. Indeed, in 1923 L. de Broglie was the first \cite{debroglie} to exploit Lorentz invariance in order to formulate the well-known relations between the energy/momentum of a particle and the frequency/wave-vector of the associated wave. According to P.A.M. Dirac \cite{dirac}, even the subsequent formal development of de Broglie's ideas did not led E. Schr\"odinger \cite{schrodinger} to write down first his most famous (non-relativistic) equation but, rather, the {\it relativistic} wave equation now known after O. Klein and W. Gordon \cite{KG} (which, for some time, was referred to as the relativistic Schr\"odinger equation). The original reasoning by Schr\"odinger started from the relativistic energy-momentum relation for an electron,
\be \label{1}
- \, \frac{W^2}{c^2} + p^2 + m^2 c^2 = 0, 
\ee
and then assumed that the electron would be described by a wave-function $\psi(\xbvec, t)$ satisfying the equation obtained by making in (\ref{1}) the replacements
\be \label{2}
\frac{W}{c} \rt \frac{i \hbar}{c} \, \frac{\prt}{\prt  t} ,  \qquad
\bvec{p} \rt - i \hbar \gr ,
\ee
from which the de Broglie's relations came out for a plane wave $\erm^{i (\bvec{k} \cdot \bvec{x} - \omega t)}$. The resulting, first discovered, wave equation was then the relativistic Klein-Gordon equation:
\be \label{KGE}
\left( \Box + \frac{m^2 c^2}{\hbar^2} \right) \psi = 0.
\ee
Schr\"odinger then abandoned his relativistic wave equation since it gave the wrong predictions for the fine structure of the hydrogen atom, but later realized that the non-relativistic approximation to his relativistic equation, the proper Schr\"odinger equation, led to some correct results despite the original relativistic formulation. Schr\"odinger and others soon recognized that the source of discrepancy between the relativistic wave equation and observations was the neglect of the spin of the electron (the Klein-Gordon equation describes spin 0 particles) but, as well-known, we have to wait until 1928 when Dirac discovered \cite{dirac1928} how to incorporate the spin of the electron in wave mechanics, in a consistent, relativistically invariant manner. 

The enormous success of the Dirac theory after the discovery of the positron in 1932 \cite{anderson}, and especially its quantum field theory formulation and incorporation in quantum electrodynamics, almost entirely obscured other subsequent formulations and generalizations of relativistic wave equations for particles with different value of the spin. Indeed, the early history of such equations is quite rich and intriguing, and extends over almost two decades. In the present paper we give (in the following section) an historical account of the different formulations of relativistic wave equations appeared in the literature till the end of 1940s, together with the physical motivations for them. Subsequent elaborations were mainly aimed at achieving mathematical improvements, and will not be considered here. It is mandatory to recall that relativistic wave mechanics derives its physical relevance just from its incorporation into quantum field theory (sometimes referred to as ``second quantization"): as later realized, indeed, a relativistic quantum theory of a fixed number of particles is an impossibility. While some mentions about the quantum field theory justification of the equations proposed will be added occasionally, given their relevance in the subsequent developments, we will not dwell upon this topic which is far beyond our aim (for several aspects, see, instead, Ref. \cite{EspPW}). We will refer chiefly to the original papers (primary sources), since no exhaustive historical reviews are known; see however the partial but beautiful, though dated, review in Ref. \cite{nieto} (see also \cite{fradkin}). While such papers tell us the known, though not widespread, story on this subject, in Sect. 3 we focus on several important results achieved by the Italian physicist Ettore Majorana \cite{EM1} \cite{EM2} but {\it not} published by him. They are contained in some booklets with his personal research notes, which only recently have been published \cite{volumetti} \cite{quaderni}, and clearly testify (once more \cite{adp}) how he anticipated key results from different authors obtained over the subsequent years. In this respect, it is as well interesting to point out how results published by Majorana were received by his contemporaries, since they were fully understood only years later (a first illuminating example being just the mentioned Ref. \cite{fradkin}). From the correspondence of W. Pauli, then, we will be able in Sect. 4 to 
directly follow the study of the seminal paper published by Majorana in 1932 \cite{EM1932}, that occupied Pauli for about one year. Finally, in Sect. 5 we summarize the results reviewed and give our conclusions.

\section{Relativistic wave equations: the known story}

\noindent The original rejection by Schr\"odinger (and others) of Eq. (\ref{KGE}) as the correct quantum equation describing an electron was based, as recalled above, on the comparison of its prediction with the accurate experimental spectroscopic data on the hydrogen atom, and such discrepancy was correctly attributed to the non consideration of the spin of the electron. However, when Dirac approached the problem of making a relativistic theory of the spinning electron, no reference to such discrepancy was made, but rather he focused on another {\it theoretical} problem, namely that of negative probabilities.

\subsection{Dirac equation for a spin-1/2 electron (1928)}

\noindent The probability density for the non-relativistic Schr\"odinger equation was known \cite{prob} to be a positive definite quantity, $\rho = |\psi|^2$, satisfying a continuity equation which renders the space integral of $\rho$ to be time-independent. On the other hand, the only $\rho$ which can be formed from the solutions of the Klein-Gordon equation and satisfying again a continuity equation does not have a definite sign, so that it is not possible to identify it as the probability density in the corresponding relativistic case. This was the major problem which Dirac attempted to solve in order to obtain a consistent relativistic theory. He then realized that the possible negative probabilities arising from the Klein-Gordon equation were due to the presence, contrary to what happened in the non-relativistic case, of a time derivative of the wave function in the expression for $\rho$, and this directly came out from the fact that the Klein-Gordon equation was a differential equation of the {\it second order} in the time variable. The dynamical evolution of the wave function should then be ruled by a first order in time differential equation, just as for the 
Schr\"odinger equation. In order to have a Lorentz-invariant theory, however, this led Dirac to assume that the relativistic wave equation to be found should be as well linear in space derivatives:
\be \label{DE}
\left( \frac{W}{c} + \alphavec \cdot \pbvec + \alpha_4 \, m c \right) \psi = 0 \, ,
\ee
where the replacements in (\ref{2}) apply. ``The $\alpha$'s are new dynamical variables which it is necessary to introduce in order to satisfy the conditions of the problem. They may be regarded as describing some internal motion of the electron, which for most purposes may be taken the spin of the electron postulated in previous theories." \cite{dirac1928} The ``condition" to be satisfied is that the energy and momentum of the particle verify the energy-momentum relation in Eq. (\ref{1}) or, through (\ref{2}), that the solution of the Dirac equation (\ref{DE}) satisfies also the Klein-Gordon equation (\ref{KGE}). Dirac showed that this was indeed the case provided that the ``new dynamical variables" satisfy the relations:
\be \label{diracalgebra} 
\ba{l}
\dis \alpha_\mu \alpha_\nu + \alpha_\nu \alpha_\mu  = 0 \quad (\mu \neq \nu),  \\
\dis \alpha_\mu^2 = 1
\ea
\ee
($\mu,\nu=1,2,3,4$). He noted that, for these relations to be true, the $\alpha$ quantities must be $4 \times 4$ matrices obtained from the Pauli $2 \times 2$ matrices $\sigmavec$,
\be \label{diracmatrices}
\alphavec = \left( \ba{cc} 0 & \sigmavec \\ \sigmavec & 0 \ea \right) \! , \qquad \alpha_4 = \left( \ba{cc} I_2 & 0 \\ 0 & I_2 \ea \right)
\ee
(where $0$ and $I_2$ are the $2 \times 2$ zero and identity matrices, respectively), so that the wave function itself is what later will be named a 4-component spinor. Dirac then proved the full Lorentz invariance of his theory: Eq. (\ref{DE}) could be written in the form
\be \label{DE2}
\left( i\gamma^\mu p_\mu + m c \right) \psi = 0 ,
\ee
with the matrices $\gamma_\mu$ transforming as a 4-vector (just as the 4-momentum $p_\mu$), and satisfying the relation:
\be \label{gammaalgebra} 
\gamma_\mu \gamma_\nu + \gamma_\nu \gamma_\mu  = 2 \delta_{\mu \nu} 
\ee
(which relation is, as well, relativistically invariant). Moreover, he obtained the hamiltonian for an electron in an arbitrary electromagentic field and, from the conservation of angular momentum, showed that ``his" electron actually had a value of the spin equal to $1/2$. Finally, he also pointed out that his theory gave correct predictions (at order $\alpha^4 m c^2$, where $\alpha$ is the fine structure constant) for the fine structure splitting of the hydrogen energy levels, contrary to what happened for the Klein-Gordon equation. However, as already pointed out, Dirac's primary aim was that of a relativistic formalism with positive probabilities, and this was actually achieved, since the probability density satisfying the continuity equation was now just given by $\rho = |\psi|^2$.

\subsection{Negative energy states}

\noindent A second major difficulty of the Klein-Gordon equation, as pointed out by Dirac himself, concerned again the fact that the energy $W$ in Eq. (\ref{1}) appears to the second power. As a matter of fact, the Klein-Gordon equation...
\begin{quote}
...refers equally well to an electron with charge $e$ as to one with charge $-e$. If one considers for definiteness the limiting case of large quantum numbers one would find that some of the solutions of the wave equation are wave packets moving in the way a particle of charge $-e$ would move on the classical theory, while others are wave packets moving in the way a particle of charge $e$ would move classically. For this second class of solutions $W$ has a negative value. One gets over the difficulty on the classical theory by arbitrarily excluding those solutions that have a negative $W$. One cannot do this on the quantum theory, since in general a perturbation will cause transitions from states with $W$ positive to states with $W$ negative. Such a transition would appear experimentally as the electron suddenly changing its charge from $-e$ to $e$, a phenomenon which has not been observed \cite{dirac1928}.
\end{quote}
This second difficulty was not removed in the original Dirac theory, as explicitly admitted in \cite{dirac1928}, and the subsequent three years saw Dirac working to solve this problem. Indeed, the quotation above should not be read -- anachronistically -- as an anticipation of the final interpretation of the negative energy states in terms of positrons, that came through a tortuous path only some years later and culminated in the experimental observations by C.D. Anderson in 1932 \cite{anderson}.

W. Heisenberg and Pauli, for example, simply discarded such a problem, regarding the negative energy states as an ``inconsistency of the theory [...], which must be accepted as long as the Dirac difficulty is unexplained'' \cite{HP1929a}. Dirac, instead, soon realized that from his equation it follows that a negative energy electron moves in an external field as if it had positive charge and energy. This original observation led H. Weyl in 1929 to ``expect that, among the two pairs of components of the Dirac [wavefunction]  one pair corresponds to the electron, while the other to the proton'' \cite{Weyl1929}. This association was further considered (for some time) by Dirac in the framework of his ``hole theory'', with the identification
of ``the holes in the distribution of negative energy electrons'' with the protons \cite{Dirac1929}, although novel difficulties arose with such interpretation. Here we only quote from the Dirac paper appeared at the very beginning of 1930: %
\begin{quote}
Can the present theory account for the great dissymmetry between
electrons and protons, which manifest itself through their
different masses and the power of protons to combine to form
heavier atomic nuclei? It is evident that the theory gives, to a
large extent, symmetry between electrons and protons [...] The
symmetry is not, however, mathematically perfect when one takes
interaction between the electrons into account. [...] The
consequences of this dissymmetry are not very easy to calculate on
relativistic lines, but we may hope it will lead eventually to an
explanation of the different masses of proton and electron
\cite{Dirac1929}.
\end{quote}
Such difficulty in the interpretation of the theory, and some others related to it (i.e., the too high rate of
annihilation of electrons and protons), were later overcome by Dirac himself, after the Weyl proof that holes necessarily represented particles with the same mass as an electron \cite{Weyl1931} \cite{Dirac1931}. This novel interpretation, though highly controversial, was eventually confirmed by the cosmic-rays experiments by Anderson \cite{anderson} in 1932. However, the fact that the observed positive electrons were indeed Dirac antiparticles was not clear for some time, but was fully recognized only after the appearance of the experimental results on cosmic-rays showers by P.M. Blackett and G.P.S. Occhialini \cite{Blackett}, followed by a discussion of their results within the framework of the Dirac theory. Notwithstanding this, especially Pauli remained very critical with the Dirac theory of
negative energy states,\footnote{``I do not believe in your perception of `holes', even if the existence of the `antielectron' is proved'', wrote Pauli to Dirac on May 1, 1933 \cite{Pauli1985}. } a position shared by another theoretician who came into play with a more general (and more difficult to understand) theory.

\subsection{Majorana infinite-component equation (1932)}

\noindent With the exception of Pauli and very few other people who gave some (justified) importance to the theoretical problem of the negative energy states, the vast majority of physicists granted the success of the Dirac equation even before its ``final" confirmation with the discovery of the positron. The reason was mainly the successful predictions about the fine structure of the hydrogen and the correct account for the magnetic moment of the electron (with the prediction of its gyromagnetic ratio $g=2$), which were a major trouble for the just born quantum mechanics. People realized that such problems were intimately related to the spin of the electron, and the successful Dirac theory offered a reasonable and attractive way out, even well founded on solid formal basis. Even better, quite a general consensus prevailed that the 1/2 value of the electron spin was a necessary consequence of the theory of relativity, i.e. of Lorentz invariance \cite{gentile}. Just the contrary was demonstrated in 1932 by Majorana \cite{EM1932} who explicitly built a consistent relativistic quantum theory for particles with {\it arbitrary} spin.

The starting point in the Majorana's paper, however, was again the persisting problem of the negative energy states, a lucid examination of which directly led Majorana to generalize the Dirac equation to particles with arbitrary spin. The relativistic wave equation was assumed to be of the same form as in Eq. (\ref{DE}), i.e.
\be \label{ME}
\left[ \frac{W}{c} + \alphavec \cdot \pbvec - \beta \, m c \right] \psi = 0 \, ,
\ee
but nothing was assumed for the $\alphavec, \beta$ matrices appearing in it. Instead, a thorough (mathematical) inspection about the insorgence of the negative energy states was presented.
\begin{quote}
Equations of this kind present a difficulty in principle. Indeed, the operator $\beta$ has
to transform as the time component of a 4-vector, and thus $\beta$ cannot be simply a 
multiple of the unit matrix, but must have at least two different eigenvalues, say $\beta_1$ and
$\beta_2$. However, this implies that the energy of the particle at rest, obtained from Eq. (\ref{ME}) by
taking $p = 0$, shall have at least two different values, i.e. $\beta_1 \, m c^2$ and $\beta_2 \, m c^2$. According to Dirac's equations, the allowed values of the mass at rest are, as well known, $+m$ and
$-m$; from this it follows by relativistic invariance that for each value of $p$ the energy can
acquire two values differing in sign:  $W = \pm \sqrt{m^2 c^4 + c^2 p^2}$.
As a matter of fact, the indeterminacy in the sign of the energy can be eliminated by using equations of the type (\ref{ME}), only if the wave function has infinitely many components that cannot be split into finite tensors or spinors \cite{EM1932}.
\end{quote}
Thus, Majorana did not require the energy-momentum relation (\ref{1}) to be satisfied for each component of the $\psi$, as instead Dirac did in order to determine the expressions for the matrices $\alphavec, \beta$. Rather, Majorana determined {\it a priori} a representation for the infinite matrices corresponding to the six infinitesimal Lorentz transformations, that is the transformation matrices for the components $\psi_{jm}$ of the wave function forming a basis for that representation. The form of the $\alphavec, \beta$ matrices was then deduced  just by imposing relativistic invariance on the action integral. The rest energy of the particles described by such theory is, as anticipated, positive:
\be  \label{rest}
W_0 = \frac{m c^2}{j+ \frac{1}{2}} \, ,
\ee
but this reveals that Eq. (\ref{ME}) is a {\it multi-mass} equation. ``For half-integer values of $j$ we thus obtain states corresponding to the values $m,m/2,m/3,\dots$ of the mass, while for integer $j$ one has $2m, 2m/3, 2m/5, \dots$. It should be emphasized that particles having different masses also have different
intrinsic angular momentum." Also, solutions with different energy eigenvalues are as well present in the theory: ``Besides the states pertinent to positive values of the mass, there are other states for
which the energy is related to the momentum by a relation of the following type: $W = \pm \sqrt{c^2 p^2 - k^2 c^4}$; such states exist for all positive values of $k$ but only for $p\geq kc$, and can be regarded as
pertaining to the imaginary value $ik$ of the mass." Tachionic solutions thus enter for the first time in a relativistic wave equation.

All these peculiarities were rediscovered and appreciated only years later (see below): the physical and mathematical content of the Majorana theory was, evidently, too much ahead of his time, during a period when people became convinced that spin 1/2 particles were enough for Nature, and that such a result directly followed from Lorentz invariance through the simple Dirac equation. The situation didn't change so  much even when, in 1934, Pauli and V.F. Weisskopf \cite{PW} succeeded in ``solving" the problem of the negative energy states for the Klein-Gordon equation (namely, by ``second-quantizing" this theory), thus showing that the marriage between quantum mechanics and special relativity did not necessarily require a spin 1/2 for the correct interpretation of the formalism, as erroneously believed.  Thus, non intrinsically wrong arguments were proved to lie in the Klein-Gordon theory that justified the derivation of the Dirac theory; such theories just applied to particles with  different spin, as already deduced by Majorana himself \cite{EspPW}.

The mathematical relevance of the Majorana's paper should not be underestimated as well. Remarkably, he indeed obtained the simplest infinite-dimensional unitary representations of the Lorentz group that were re-discovered by E.P. Wigner in his 1939 and 1948 works \cite{wigner}. It is quite interesting also to dwell on the extremely simple reasoning which led Majorana to consider the infinite-dimensional representations, easily recognizable in his personal notebooks (see page 446 in Ref. \cite{volumetti}):
\begin{quote}
The representations of the Lorentz group are, except for the identity representation, essentially not unitary, i.e., they cannot be converted into unitary representations by some transformation. The reason for this is that the Lorentz group is an open group. However, in contrast to what happens for closed groups, open groups may have irreducible representations (even unitary) in infinite dimensions. In what follows, we shall give two classes of such representations for the Lorentz group, each of them composed of a continuous infinity of unitary representations.
\end{quote}
As well interesting is the acknowledgment by Wigner of the Majorana's work, typical of a rigorous mathematician:
\begin{quote}
The representation of the Lorentz group have been investigated repeatedly. The first investigation is due to Majorana, who in fact found all representations of the class to be dealt with in the present work excepting two sets of representations. [...] The difference between the present paper and that of Majorana [...] lies -- apart from the finding of new representations -- mainly in its greater mathematical rigor. Majorana [...] freely uses the notion of infinitesimal operators and a set of functions to all members of which every infinitesimal operator can be applied. This procedure cannot be mathematically justified at present, and no such assumption will be used in the present paper. Also the conditions of reducibility and irreducibility could be, in general, somewhat more complicated than assumed by Majorana \cite{wigner}.
\end{quote}
The greater mathematical rigor by Wigner certainly justifies the 56 pages of his paper, but the answer-at-a-distance by the physicist Majorana is illuminating: ``in order to avoid exaggerated complications, we will give the transformation law only for infinitesimal Lorentz transformations, since any finite transformation can be obtained by integration of the former ones." Such an approach is still used by practically any physicist.

Despite such simplicity of reasoning, this Majorana theory was apparently not appreciated by many, given its very few traces in the literature:\footnote{This is related also to the fact that it was published in a not internationally renowned journal. As already noted by Fradkin \cite{fradkin}, ``Science Abstracts (Section A, Physics), the English language abstracting service, did not abstract from {\it Nuovo Cimento} until 1946. Majorana's article was given a several line abstract by the contemporary German abstract service [Physikalische  Berichte, 1933-I, p. 548] but the abstractor, whose major field was fluorescence of salts and crystal studies, failed to assess its significance or even mention the occurrence of the infinite-dimensional representations.} apart from Wigner, for what we are concerned with here it was quoted only in \cite{kemmer} \cite{gentile} \cite{lubanski} for the period of time considered. Nevertheless, this is only partially true (especially for Pauli, see below), since much of the ideas underlying it and even results obtained  by him will be rediscovered and reconsidered later by other people in different times, as we will see in the following.

\subsection{Majorana-Oppenheimer formulation of electrodynamics (1931)}

\noindent After the success of the first-order Dirac equation in describing a relativistic quantum particle, a quite peculiar belief made its way that the ``wave" properties of a (neutral) field be described by a (second-order) ``wave" equation as in the Klein-Gordon case, while the ``particle" properties be instead described by a (first-order) ``particle" equation as in the Dirac case. While this belief was latent in the minds of physicists in the 1930s, and the question was properly addressed only with the clear emergence of quantum field theory, the words by N. Kemmer still in 1939 are unambiguous:
\begin{quote}
\indent In the case of particles with an electric charge [...] the transition to the limiting theory of a classical {\it particle} is possible. 

On the other hand, for any uncharged particle [...] a limiting classical particle theory does {\it not} exist. 

Conversely, one finds that, at least in the Bose case, a complete correspondence of the theory of uncharged particles with a classical {\it wave} theory exists, whereas the correspondance appears to {\it fail} for charged particles. The latter fact can well be understood, for a classical entity corresponding to the quantum mechanical ``charged field" is hard to envisage.

For the special case of the {\it meson} we thus have the following peculiar situation: although it seems likely that both charged and uncharged mesons exist, and the quantum treatment of the two is well nigh identical, the uncharged one [...] is classically a true field, the uncharged one, on the other hand, a particle \cite{kemmer}.
\end{quote}
The quite obvious conclusion was that such a situation ``is neither justified by experimental considerations nor by arguments of correspondance."

The Kemmer's own solution of this issue, initially developed for the particular case of mesons, will be described below, while we now dwell a bit on the earlier consideration of the problem.

The first ``particle" for which it was noted such an asymmetric description was the photon. It is described as a ``classical" field by the second order D'Alembert wave equation deduced by the Maxwell equations, but it was J.R. Oppenheimer \cite{oppy} who first considered the problem of finding a first-order ``particle" equation for it: ``our present problem is to find the wave equation for the de Broglie waves of the [light] quantum." The ``equation of the retarded potentials" 
\be
\Box \psi =0, 
\ee
indeed, according to Oppenheimer is ``in several respects unsatisfactory. Here, just as for the electron, we should want a linear equation, in order to obtain a suitable density flux vector with vanishing divergence." In the much simpler (and independent, though not published) formulation by Majorana \cite{EMphoton}, such a problem is just to translate the (first-order) Maxwell equations into a Dirac-like form for a suitable photon wave function. Oppenheimer showed that such a wave function ``no longer behaves as an invariant under space rotations, but as a spinor of the first rank", and, in particular, a 3-component ``spinor" (that is, a vector) is required. By taking $\psivec = \Ebf- i c \Bbf$, where $\Ebf, \Bbf$ are the electric and magnetic field of the associated electromagnetic wave, the four Maxwell equations (in vacuum) can indeed be written in the form of a Dirac equation:
\be
\left( \frac{W}{c} - \alphavec \cdot \pbvec \right) \psivec = 0 \, ,
\label{MaxwellDirac}
\ee
with the necessary transversality condition $\pbvec \cdot \psivec = 0$. The mass of the particle is, of course, zero. The quantities $\alphavec$ are $3 \times 3$ hermitian matrices given by
\be
\alpha_1 \; = \;
\left( \ba{ccc}
0 & 0 & 0 \\
0 & 0 & i \\
0 & -i & 0
\ea \right), \quad
\alpha_2 \; = \;
\left( \ba{ccc}
0 & 0 & -i \\
0 & 0 & 0 \\
i & 0 & 0
\ea \right) ,\quad
\alpha_3 \; = \;
\left( \ba{ccc}
0 & i & 0 \\
-i & 0 & 0 \\
0 & 0 & 0
\ea \right) ,
\ee
which satisfy the relations 
\be \label{EMalgebra}
\left[ \alpha_i , \alpha_j \right] = - i \, \epsilon_{i j k}
\, \alpha_k \, .
\ee
The probabilistic interpretation of such a theory is indeed possible given the continuity equation satisfied by the wave function $\psivec$, that equation being just the translation into the actual language of the Poynting theorem for the electromagnetic field (see the last reference in \cite{EMphoton}). However, as recognized by Oppenheimer himself, the present theory for the photon is unsatisfactory for several reasons, the main one concerning the non-explicit Lorentz invariance, given the employment of a 3-vector instead of a covariant 4-vector. Nevertheless, the novel basic concepts introduced will be longlasting:
\begin{enumerate}
\item a directly observable wave function (related to the electric and the magnetic fields, rather than to the electromagnetic potentials) is worth of consideration;
\item the spinor space is not necessarily that introduced by Dirac, a different dimension and algebra (as given in (\ref{EMalgebra})) being possible.
\end{enumerate}
The discussion of the first point, though notably interesting, is out of what considered here, and will not be further discussed. Instead, some effort was devoted since 1931 to the second point, as already introduced in the above discussion of the Majorana infinite-component equation, and will be the subject of the following paragraphs.

\subsection{de Broglie theory of a ``composite" photon (1934)}

\noindent The starting point of the Majorana-Oppenheimer theory was mainly theoretical in nature, being inspired to the desire of ``translating" the Maxwell equations into a Dirac-like form. This point was further considered in 1934 (without reference to Oppenheimer) by de Broglie, who started from a full physical inspiration:
\begin{quote}
Nous avions propos\'e de consid\'erer le photon comme form\'e par deux corpuscoles compl\'ementaires qui soinet l'un par rapport \`a l'autre comme l'\'electron positif par rapport au n\'egatif et nous avions rattach\'e \`a cette conception une d\'efinition des champs \'electromagn\'etiques li\'es au photon \cite{deb1934}.
\end{quote}
Here, the original idea behind the theory is evidently related to the known process of electron-positron annihilation into photons, which could easily lead to the (erroneous) belief that the photon was effectively composed of two of such particles,\footnote{Note that the successful Fermi theory of nuclear beta decay \cite{fermi}, where a similar problem was correctly addressed, was contemporary to the paper by de Broglie.} with the implicit prediction of a massive photon. In order to achieve his proposal, de Broglie needed to introduce ``une \'equation des ondes pour le photon au lieu de partir, ainsi que nous l'avions fait, de l\'equation de Dirac pour le demi-photon." The de Broglie equation for the photon had just the same form as that of a Dirac equation,
\be \label{DB}
\frac{1}{c} \frac{\prt \Phi}{\prt t} = \frac{\prt }{\prt x} \, A_1 \Phi +  \frac{\prt }{\prt y} \, A_2 \Phi + 
 \frac{\prt }{\prt z} \, A_3 \Phi + \frac{2i \mu_0 c}{\hbar} \, A_4 \Phi \, ,
\ee
where $\mu_0$ is the mass of the ``semi-photon". Here the interesting novelty is the introduction of the four $16 \times 16$ matrices $A_\mu$ obtained as products of different Dirac matrix spaces (explicitly reported in \cite{deb1934}) in accordance with the original idea envisaged above. Indeed, as de Broglie himself realized, such matrices did {\it not} follow the Dirac algebra (\ref{diracalgebra}), though the component $\alpha$ matrices did. As we will see below, it was a student of de Broglie, G. Petiau, who again worked on this $16 \times 16$ matrices and became the first to discover the so-called DKP algebra \cite{petiau}. For the moment, however, we only mention an interesting property of one possible solution of the de Broglie equation, namely the solution which ``correspond \`a une \'energie, une quantit\'e de mouvement et un spin nuls. Elle est donc appropri\'ee \`a la repr\'esentation de l'\'etat d'annihilation du photon."

de Broglie considered his hypothesis (the so-called "neutrino theory of the photon") for more than 40 years \cite{nieto}, despite the experimental evidence against it (the basic prediction is a non-vanishing mass for the photon which, however, was not observed). 

\subsection{Dirac-Fierz-Pauli generalized equations for spin higher than 1/2 (1936-9)}

\noindent The appearance of the Dirac equation gave for the first time an example to physicists of a system of equations which is invariant in form when subjected to  a Lorentz transformation, but which is not written in terms of tensors. Indeed, though not supported by mathematicians, physicists tacitly assumed that the ordinary tensor language did comprise all possible representations of the Lorentz group, while Eq. (\ref{DE2}) explicitly showed this is not true. Spinor calculus was worked out basically by B. van der Waerden \cite{van}, who also showed how write the Dirac equation in an automatically covariant form. Though particularly difficult for the physicists of the time uninitiated to this mathematical machinery, such a powerful instrument was completely adopted by Dirac (and few others) who succeeded in writing down a generalization of his earlier equation to particles with integral or half-odd integral spin greater than 1/2. It connects two irreducible spinors; for a particle of spin $n + 1/2$ and mass $m$; in spinor notation it was written as follows:
\be  \label{DFP}
\ba{l}
\dis p_{\gamma \dot{\alpha}} A_{\epsilon_1 \epsilon_2 \dots \epsilon_n}^{\dot{\alpha} \dot{\beta}_1 \dot{\beta}_2 \dots \dot{\beta}_n} = m c \, B_{\gamma \epsilon_1 \epsilon_2 \dots \epsilon_n}^{\dot{\beta}_1 \dot{\beta}_2 \dots \dot{\beta}_n} , \\ \\
\dis p^{\gamma \dot{\alpha}} B_{\gamma \epsilon_1 \epsilon_2 \dots \epsilon_n}^{\dot{\beta}_1 \dot{\beta}_2 \dots \dot{\beta}_n} = m c \, A_{\epsilon_1 \epsilon_2 \dots \epsilon_n}^{\dot{\alpha} \dot{\beta}_1 \dot{\beta}_2 \dots \dot{\beta}_n} .
\ea
\ee
The component wave functions $A$ and $B$ are each completely symmetric in their dotted and undotted spinor indices, and  $p_{\gamma \dot{\alpha}}$ is the momentum operator written as a covariant spinor. For $n=0$, Eqs. (\ref{DFP}) are equivalent to the Dirac spin-1/2 equation and the two spinors $A^{\dot{\alpha}}$, $B_\gamma$ transform like a Dirac bi-spinor.\footnote{A spinor with one dotted and one undotted index transforms like a 4-vector \cite{van} so that, by pairing $\dot{\beta}_1 $ and $\epsilon_1$, one could replace $n$ dotted and $n$ undotted spinor indices of both $A$ and $B$ by a $n$ symmetric traceless 4-vector indices. Therefore, Eqs. (\ref{DFP}) could be written in the same formalism as in Eq. (\ref{DE}) with a wave function having, beside the bi-spinor index, $n$ symmetric traceless 4-vector indices:
\be  \label{DFP2}
\left( \frac{W}{c} \, \delta^{ab}+ \alpha_i^{ab} p_i + \alpha_4^{ab} \, m c \right) \psi^b_{\nu_1 \nu_2 \dots \nu_n} = 0 \, .
\ee
However, since the powerful mathematical formalism offered by spinor calculus, though somewhat more involved, plays not at all a secondary role in the theory originallly proposed (as well as in subsequent re-considerations), we prefer to draw the reader's attention to Eqs. (\ref{DFP}) rather than to (\ref{DFP2}).} The basic requirement considered by Dirac that led to Eqs. (\ref{DFP}) is that each component of the wave function sarisfies the Klein-Gordon equation (\ref{KGE}), just as in the Dirac spin-1/2 equation. Indeed, either $A$ or $B$ can be eliminated from the equations, thus remaining with a second-order equation (\ref{KGE}) satisfied by each component separately, {\it provided that} the following condition is applied:
\be \label{subs}
p^\gamma_{\dot{\alpha}} B^{\dot{\alpha} \dot{\beta}_2 \dots \dot{\beta}_n}_{\gamma \epsilon_1 \epsilon_2 \dots \epsilon_n} = 0
\ee
(and similarly for $A$). Actually, such a condition effectively holds, since it can be deduced, for instance, by contracting the indices $\dot{\alpha}$ and $\dot{\beta}_1$ in the second of Eqs. (\ref{DFP}), so that the original aim by Dirac was fulfilled.

Apart from the finite number of components considered in the Dirac theory, the requirement that Eq. (\ref{KGE}) holds for any component of the wave function is a remarkable difference with respect to the Majorana theory of 1932, since it is physically equivalent to require that the particle described by the wave function have (in each case) only one value of the mass -- except for the sign --, contrary to what happened in the Majorana theory. The same is true for the spin: each equation describes a particle with only one spin state.

The problems with the negative energy states were solved by M. Fierz in 1939 \cite{fierz} just in the same way as for the Dirac spin 1/2 equation and for the Klein-Gordon equation, that is by setting up a scheme of second-quantization for the theory (in the absence of an external field). This led Pauli to consider (again with Fierz) the obvious subsequent case of second quantization when an external electromagnetic field is instead applied \cite{fp}, but here a surprising fact came out even without considering at all field quantization. In fact, Fierz and Pauli discovered that ``the most immediate method of taking into account the effect of the electromagnetic field, proposed by Dirac (1936), leads to inconsistent equations as soon as the spin is greater than 1."

The interaction with an electromagnetic field was usually introduced by replacing the rules in (\ref{2}) with the standard ones in the minimal coupling scheme:
\be \label{mc}
\frac{W}{c} \rt \frac{i \hbar}{c} \, \frac{\prt}{\prt  t} + e \varphi ,  \qquad
\bvec{p} \rt - i \hbar \gr + e \bvec{A},
\ee
where $(\varphi, \bvec{A})$ are the electromagnetic potentials. The problem pointed out by Fierz and Pauli 
derived from the {\it subsidiary condition} in Eq. (\ref{subs}). Indeed, while the equations in (\ref{DFP}) can be deduced from a variational principle (that is, a lagrangian function may be written leading to those equations, as in the Dirac 1/2 case or in the Klein-Gordon case or even in the Majorana case), from what pointed out above it is clear that this does not happen for the subsidiary condition (\ref{subs}), since it is derived directly from the equations of motion (\ref{DFP}). This was not a really relevant point for the field-free case, but the contrary is true when the replacements (\ref{mc}) were applied directly to Eq. (\ref{subs}) (and not in the lagrangian function), since the spinor fields should satisfy additional constraints that ``cannot in general be satisfied simultaneously with the other equations." Fierz and Pauli solved the problem by deducing a lagrangian function from which both Eqs. (\ref{DFP}) and (\ref{subs}) can be obtained by a rather complicated procedure. ``This consists in introducing auxiliary tensors or spinors of lower rank than the original ones [...] and deriving all equations from a variation principle without having to introduce extra conditions" \cite{fp}. The arbitrary constants present in the lagrangian function were adjusted in a way that, in the free case, the equations of motion led to the identical vanishing of the auxiliary tensors/spinors, while Eqs. (\ref{DFP}) and (\ref{subs}) held true, and the interaction with an electromagnetic field could then be introduced by the standard procedure in (\ref{mc}). The ``artifice" employed by Fierz and Pauli, however, was not very elegant since, in the presence of interaction, the auxiliary spinors did not vanish identically, though there were just so many equations that for any given value of the momentum of the particle there were the right number of independent states corresponding to the spin of the particle at hand. Thus, the procedure worked fine, but became progressively more complicated for larger spins.

\subsection{Proca equation for massive spin-1 particles (1936)}

\noindent The de Broglie's idea of a composite photon influenced for some time physicists in the 1930s, especially those who worked in France. This was the case of A. Proca, a Romanian theoretician whose doctoral advisor was de Broglie himself. In 1936  \cite{procac} he interpreted the two (complex conjugated) terms in the plane wave expansion of the photon field in quantum electrodynamics as the ``superposition" of two elementary particles (as happened for the quantized Dirac electron-positron field) following the de Broglie original idea:
\begin{quote}
ces particules ont m\^eme \'energie ({\it positive}), m\^eme quantit\'e de mouvement et m\^eme spin, mais des charges, es courants et des moments \'electromagn\'etiques oppos\'es. L'important est le fait qu'{\it elles ne sont plus des neutrinos}, mais des particules {\it charg\'ees}, de masse nulle; ce sont en quelque sorte des {\it charges pures} \cite{procac}.
\end{quote}
Thus, according to Proca \cite{proca}, ``la th\'eorie de la lumi\`ere entre par cette voie dans le cadre dÕune m\'ecanique quantique g\'en\'erale qui englobe notamment les \'electrons n\'egatifs et les positons.'' The basic concept he introduced was that of ``pure charge'' particles, with a zero rest mass (roughly speaking, massless electrons and positrons),  which were assumed to be the two components of the photon. This served to Proca to describe the photon by means of a 4-component wave function $\psi_r$ (where, however, $r$ is a Lorentz index), each component of which being treated as independent scalar wave functions. Here, although Proca always took the limit of zero mass for the application to the photon, he nevertheless considered the general case of a massive spin-1 particle. In this case, each component had to satisfy the Klein-Gordon equation (for a free particle):
\be \label{proca1}
\Box \psi_r  = k^2 \, \psi_r 
\ee
($k=mc/\hbar$). However, Proca showed that, in order to have a positive energy (and, more in general, a positive definite energy density at every space point), the supplementary condition (and its complex conjugate)
\be \label{procasup}
\prt^r \psi_r = 0
\ee
must hold. 

A meaningful reformulation in terms of first-order rather than second-order differential equations (in analogy to the ``translation'' of the Klein-Gordon equation into the Dirac equation) was as well given. Indeed, inspired by Maxwell electrodynamics, a skew-symmetric tensor 
\be \label{skew}
G_{rs} =  \prt_r \psi_s - \prt_s \psi_r
\ee
(and its conjugate) was introduced, the equation of motion for which are just generalizations of the Maxwell equations to non-zero photon mass:
\be \label{proca2}
\prt^r G_{rs} = k^2 \, \psi_s .
\ee
From these, it immediately followed that the equations satisfied by the wave components $\psi_r$ wrote as
\be \label{proca3}
\Box \psi_r - \prt_r \left( \prt^s \psi_s \right) = k^2 \, \psi_r , 
\ee
but from (\ref{proca2}) it followed as well, by simply using the skew symmetry of $G_{rs}$, that the condition (\ref{procasup}) holds, so that Eq. (\ref{proca1}) was recovered.

The Proca equations (\ref{proca2}) (or the equivalent forms (\ref{proca1}), (\ref{procasup})) can be obtained from a variational principle with an appropriate lagrangian function that is a thorough generalization of the Maxwell electrodynamic lagrangian, so that the introduction of the interaction with an external field is straightforward by means of the usual minimal coupling principle. Here, however, the very interesting thing, as pointed out clear by Pauli few years later, was that gauge arbitrariness for the field $\psi_r$ no longer exists, contrary to the case for the massless photon described by the 4-potential $A_\mu$: $\psi_r$ ``is uniquely defined by a given [$G_{rs}$], just as [$G_{rs}$] is defined by [$\psi_r$] from [(\ref{skew})]. As a consequence, for non-vanishing rest mass, the addition of a gradient to [$\psi_r$] is not permitted. Hence no gauge transformations of the second kind exist for the [$\psi_r$]'' when $m\neq0$ \cite{Pauli1941}.

\subsection{Majorana ``symmetric'' equation for spin-1/2 fermions (1937)}

\noindent In 1937 Majorana published a paper \cite{EM1937} containing a theory already elaborated some years earlier \cite{quaderni}. Here the main aim was not that of writing down a novel equation, but rather that of reformulating the existing Dirac equation (\ref{DE}) in order to achieve a complete ``symmetry'' between the electron and positron components described by it. Already in 1933 Heisenberg noted the substantial symmetry, in the Dirac theory, between processes involving electrons and those involving positrons \cite{20HT}, and even some time earlier Heisenberg himself elaborated an interesting application in which he considered  the symmetry between holes and electrons in an occupied atomic level or in an occupied energy band of a crystal \cite{22HT}. However, such a general idea of a particle-antiparticle symmetry was formally developed into a consistent theory only by Majorana in his most famous article on {\it a symmetric theory of electrons and positrons} \cite{EM1937}.
The equation considered by him is just the Dirac equation written in the form:
\be \label{MEST}
\left[ \frac{1}{c} \frac{\prt ~}{\prt t} - \alphabf \cdot \grbf - i \beta \, \frac{m c}{\hbar} \right] \psi = 0 ,
\ee
but here Majorana introduced a choice for the four independent $\alphabf, \beta$ matrices {\it different} from the standard Dirac's one, namely
\be \label{MEmat}
\alpha_x = \sigma_x \otimes \sigma_x, \quad
\alpha_y = \sigma_z \otimes 1, \quad
\alpha_z = \sigma_x \otimes \sigma_z, \quad
\beta = -\sigma_x \otimes \sigma_y, 
\ee
where $\sigmabf=(\sigma_x,\sigma_y,\sigma_z)$ are Pauli matrices. Such a choice led to profound implications on the theory, since $\alpha_x, \alpha_y, \alpha_z$ and $- i \beta$ all have only real elements, so that Eq. (\ref{MEST}) is an equation with real-valued coefficients. The bispinor field $\psi$ can, of course, again be decomposed into a real and imaginary part, $\psi =U+i V$, as in the Dirac theory, but now the separate equations for $U$ and $V$ are completely {\it identical}. Thus, in the quantum description of charged particles, the present theory was completely symmetric with respect to particles and antiparticles. Majorana was aware of the fact that such an advantage was purely formal, since there was no distinction between the two theories in physical applications (but with the important result that the cancellations of infinite constants, relative to single field modes, is {\it required} by the symmetrization of the theory). However, Eqs. (\ref{MEST}), (\ref{MEmat}) had in addition a {\it different} solution not present in the Dirac theory, that is a real solution $\psi = U$, without the introduction of the $V$ field. In this case, the theory described a chargeless particle or, rather, according to Majorana's own words, Eqs. (\ref{MEST}), (\ref{MEmat}) constituted ``the simplest theoretical representation of neutral particles", without the need of antiparticles. Majorana then provided also the necessary formal developments aimed at giving a solid field-theoretic basis to Eqs. (\ref{MEST}), (\ref{MEmat}) , which were derived from a variational principle by means of an appropriate lagrangian function (containing {\it only} the $U$ field). As showed earlier by Dirac and Pauli-Weisskopf, this was not a trivial task, but Majorana, guided just by mathematical elegance and symmetry, succeeded to make the idea that spin-1/2 particles could be their own antiparticles theoretically respectable, that is, consistent with the general principles of relativity and quantum theory, as already known for the photons. This was acknowledged by several people just after the appearance of Majorana's paper, including Pauli, who appreciated both ``the procedure of Majorana'' \cite{fp} (that is, the field-theoretic derivation) and the ``decomposition with respect to charge conjugate functions" \cite{Pauli1941}.\footnote{Other people included Kemmer \cite{kemmer}, Wigner \cite{wigner} and F.J. Belinfante \cite{belinfante}.} However, as in the case of other writings of his, the ``Majorana neutrino'' theory too started to gain prominence only decades later, beginning in the 1950s.

\subsection{Kemmer equation and the DKP algebra (1939)}

\noindent The successful effort to explain at least some of the features of nuclear forces made by H. Yukawa in 1935 \cite{yukawa} allowed physicists to deal seriously with novel particles different from the known spin-1/2 ones (proton, neutron, electron and positron). Indeed, Yukawa showed that the short-ranged interaction between two nucleons was mediated by a boson particle with mass intermediate  between those of the proton and the electron, which could be termed ``meson''.\footnote{As funny noticed by R. Peierls in 1939 \cite{peierls}, ``none of the properties of the new particle seems to give rise to more controversy than its name. After the names U-particle, x-particle, heavy electron, yukon, barytron, dynatron, mesotron, mesoton and meson, and, for the neutral particle, neutretto, had been used by different authors, the choice seems now to lie between mesotron and meson, of which I shall adopt the latter."} In 1938 it was definitively proved that
a particle of this type was present in cosmic rays \cite{muon}, but occurred about ten years \cite{pancini} to recognize that the meson discovered in 1937-8 was different from that hypothesized by Yukawa. Nevertheless, both the success of the Yukawa theory and the observation of mesons in cosmic rays gave quite a strong impetus to the search and study of possible equations describing mesons. At that time, no firm experimental information existed about the spin of both charged and neutral mesons, and as far as Yukawa's theory was concerned it could be either 0 or 1. Now, while equations for both spin-0 (Klein-Gordon) and spin-1 (Proca) already existed, as well as general equations in generalized spinor notation (Dirac-Fierz-Pauli), in 1939 Kemmer developed a theory based on a ``novel'' equation, whose good fortune lasted for several years. The reason is that it was becoming clear \cite{moller} that the low-energy nuclear interactions were due to pseudoscalar (spin 0) and vector (spin 1) mediators, and the Kemmer equation described them {\it both}.

The definitive formulation of the Kemmer equation has behind it an interesting historical development (reconstructed in Ref. \cite{nieto}), which involved several people working on different subjects. The story starts with the already mentioned Petiau who, in 1936 \cite{petiau}, looked at a modification of the algebra of the matrices appearing in the de Broglie equation (\ref{DB}), thus discovering the algebra satisfied by the four $16 \times 16$ matrices $\beta_\mu$ introduced:
\be  \label{DKP}
\beta_\mu \beta_\nu \beta_\lambda + \beta_\lambda \beta_\nu \beta_\mu = \beta_\mu \delta_{\nu \lambda} + \beta_\lambda \delta_{\mu \nu} .
\ee
However, Petiau's work remained practically unknown to almost everyone for many years, and the same happened to a paper by J. G\'eh\'eniau \cite{geheniau} who, two years later, decomposed such algebra into the ten-, five- and (trivial) one-dimensional representations. Meanwhile, Kemmer studied the second-order Proca equations, and found \cite{kemmer1938} that they could be described as a set of coupled first-order equations, which he wrote down along with the corresponding equations for the spin-0 case. He then realized that such equations could be written in $10 \times 10$ (for spin 1) and $5 \times 5$ (for spin 0) matrix forms, though he did not recognize the algebra satisfied by them. Some time after, the mathematician R.J. Duffin ran into the paper by Kemmer (during a seminar), and re-written \cite{duffin} both the spin-0 and the spin-1 equations into a first-order matrix formulation with the $\beta$ matrices appearing there satisfying the algebra in (\ref{DKP}) (though he did not  consider one of the four constituent commutation relations in (\ref{DKP}), namely that with $\lambda=\mu \neq \nu$). ``When Kemmer saw the note that Duffin published on his results, he wrote to Duffin saying he knew how to extend the theory but would first wait for Duffin to publish anything further. However, Duffin was by then involved in a collaboration with A.C. Schaeffer on function theory, and so wrote to Kemmer to go ahead. Kemmer then quickly put together all he had been doing, and produced his classic 1939 paper'' \cite{nieto}.

The main purpose of Kemmer was that of re-formulating the Proca equations (\ref{proca3}) in order to obtain first-order wave equations without using the tensor form in (\ref{proca2}), the theory being suitable to ``be developed in strikingly close correspondence to Dirac's electron theory''. The equation he wrote down was, then, the following:
\be  \label{KE}
\partial^\mu \beta_\mu \psi + k \, \psi = 0
\ee
($k=m c /\hbar$), where the only asumption about the four matrices $\beta_\mu$ (following Dirac's original reasoning) was that they satisfy the algebra in (\ref{DKP}). He also gave the explicit form of these matrices in the three inequivalent (ten-, five- and one-dimensional) irreducible representations of the related algebra and, probably unexpected, he discovered that Eqs. (\ref{KE}) described {\it both} spin-1 {\it and} spin-0 particles:
\begin{quote}
the ten-row representation simply leads to the usual theory based on Proca's (1936) equations, in which the wave function consists of four components forming a 4-vector and six forming an antisymmetrical tensor; the five-row one to the Klein-Gordon or so called ``scalar'' theory, in which the wave function consists of a scalar and its 4-gradient \cite{kemmer}.
\end{quote}
The present theory was, then, viewed as a generalization and a unification of previous theories and, as such, a comparison with the more general Dirac-Fierz-Pauli formalism discussed above was claimed. Kemmer realized that it could be carried out easily if an hamiltonian formulation was as well given. and succeeded in providing it by obtaining the appropriate hamiltonian function from which Eq. (\ref{KE}) can be deduced:
\be \label{KE2}
H = \frac{c \hbar}{i} \partial_k \left(\beta_k \beta_4 - \beta_4 \beta_k \right) + m c^2 \beta_4.
\ee
The conclusion was that ``Dirac's hamiltonian formulation shows clearly that the differences of his theory compared with the present one are merely due to details of representation.'' Nevertheless, Kemmer's formulation proved more simple to handle than the difficult spinor formalism and, due to the nuclear physics observations mentioned above, it enjoyed a certain consideration among physicists for some time. Starting from early 1950s, when it became clear that the observational particle situation get more involved than in the late 1930s,  physical interest in Kemmer's equation diminished: why use, for example, a ``complicated'' five-component spinor equation when we can use a one-component equation (albeit second-order) much easier to handle and leading to the same results? The situation changed a bit in 1970s (see \cite{nieto} and references therein), when the problem of the equivalence between first-order and second-order equations was considered again, finally realizing that the equivalence holds true only for free particles, while false for interacting particles.

\subsection{Further developments in 1939-1942}

\noindent After several proposals of relativistic wave equations, it was time to reason on what had been achieved and to clarify some issues. 

In this respect, a first important mathematical paper by Wigner (already mentioned) appeared in 1939 \cite{wigner}, dealing with unitary representations of the inhomogeneous Lorentz group. If the wave function
describing the possible states of a free particle satisfies a relativistic wave equation, then a correspondence exists between the wave functions describing the same state in different Lorentz frames. These transformations form the group of all inhomogeneous Lorentz transformations, and a classification of all unitary representations of the Lorentz group amounts to a classification of all possible relativistic wave equations. Wigner then gave a classification of such unitary irreducible representations, together with a prescription for their explicit construction, a remarkable result of his analysis being that every irreducible wave equation is equivalent to a system of differential equations (see, however, also the discussion above  about Majorana's work of 1932 \cite{EM1932}).

On a more physical side, it is noteworthy the general talk by Pauli at the Solvay Congress of 1939 (and published in 1941) \cite{Pauli1941}, where the theoretical physicist gave a lucid review (with further insights) of relativistic theories for spin-0, spin-1/2 and spin-1 particles, including the ``special synthesis'' of  Kemmer equation. The approach is by then field theory oriented, with a variational principle based on the lagrangian formalism as starting point. The report ended with an analysis of some physical applications of the theories discussed dealing with the interaction of particles with spin 0, 1/2 and 1 with the electromagnetic field. A special mention deserves the harsh criticism made by Pauli, already timidly expressed by Kemmer \cite{kemmer}, about the de Broglie theory of composite photons: ``on the basis of the interpretation of this paper, however, the de Broglie theory does not describe photons at all, but rather is a unified description of two particles with equal non-vanishing rest-mass, with spin values 0 and 1.''

A remarkable extension of Dirac equation to higher spin particles, in the spirit (but different formalism) of Kemmer equation, was that of F.J. Belinfante in terms of ``undors'' \cite{undors}. Belinfante assumed particles with spin $N/2$ to be described not by spinors but by Dirac wavefunctions $\psi(x, \zeta_1, \dots , \zeta_N)$ depending on $N$ 4-component variables $\zeta_r$, on which act the Dirac matrices $\gamma^\mu_r$. The wavefunction $\psi$ is assumed to be symmetric in all the  $\zeta_r$ and to satisfy Eq. (\ref{KE}) with
\be \label{beli}
\beta^\mu = \frac{1}{2} \sum_{r=1}^N \gamma^\mu_r .
\ee
Such wavefunctions, products of Dirac wavefunctions (so as to include spatial reflections) were called {\it undors} by Belinfante, who considered them as generalizations ``of Dirac wavefunctions in the same sense as tensors form a generalization of vectors.'' From the symmetry of $\psi$, however, it turned out that the Belinfante equation satisfied by undors is equivalent to $N$ identical Dirac equations with a rescaled mass (with a factor $2/N$) for each set of $\gamma^\mu_r$. The work by Belinfante was, then, a different (with respect to Dirac-Fierz-Pauli theory) generalization of the original Dirac's theory without making recourse to spinor calculus, but with the same problems envisaged above regarding the introduction of auxiliary conditions when interaction with external electromagentic fields is included.

The original Dirac-Fierz-Pauli spinor formalism was, instead, considered in 1940 by G. Gentile \cite{gentile}, who obtained a general expression for an invariant operator from which any relativistic equation for arbitrary spin, in the Dirac spinor form, could be deduced (in the Dirac-Fierz-Pauli theory, the corresponding equation was directly wrote down heuristically). In particular, he applied the result obtained to get the relativistic equation for the by then fashionable spin-1 meson\footnote{Fierz \cite{fierz} considered only implicitly such a case, while Fierz and Pauli \cite{fp} considered only the $s=2, 3/2$ cases.}:
\be \label{GentE}
\ba{l}
\dis p_{\mu \dot{\lambda}} A^{\dot{\lambda} \dot{\alpha}} = m c \, B_\mu^{\dot{\alpha}} , \\
\dis p^{\dot{\lambda} \mu} B^{\dot{\alpha}}_\mu = m c \, A^{\dot{\lambda} \dot{\alpha}} ,
\ea
\ee
with the auxiliary condition (allowing $B^{\dot{\alpha}}_\mu$ to have only three, rather than four, independent components, as for the symmetric $A^{\dot{\lambda} \dot{\alpha}}$):
\be
p^{\dot{\lambda} \mu} B^{\dot{2}}_\mu - p^{\dot{2} \mu} B^{\dot{\alpha}}_\mu = 0 .
\ee

The problem of subsidiary conditions was further investigated by the japanese physicists S. Sakata and M. Taketani in 1940 \cite{sakata}, who produced a different (and physically meaningful) formulation of the Kemmer equation. By use of what is known as a Peirce decomposition \cite{sakata}, they were able to separate out the $2(2s+1)$
(for particle and antiparticle) components of the Kemmer equation into one dinstinct hamiltonian equation. The remaining components (esssentially the built-in subsidiary conditions) were in a distinct equation that had to be satisfied simultaneously in order to obtain a covariant description. According to the Sakata-Taketani decomposition method, then, the five- and ten-component equations for spin-0 and spin-1 particles reduced to two- and six-component equations, respectively (``particle components'').

Further consideration of the Kemmer equation, as a mean to obtain a general equation for arbitrary spin, led  J.K Lubanski in 1942 \cite{luba1942} to propose the SO(5) algebra for a first-order wave equation, inspired by previous collaborations with Belinfante:
\be  \label{LubE}
\left( \Gamma_\mu \partial^\mu + N \, k \right) \psi = 0 .
\ee
The matrices $\Gamma_\mu$ were again built on from the Dirac $\gamma$ matrices, and Lubanski showed that the ten matrices given by $(i/2)\Gamma_\mu$, $(1/4) [\Gamma_\nu, \Gamma_\lambda]$ represented the operators for the infinitesimal rotations in a five-dimensional space. The quantity $N$ in (\ref{LubE}) was an integer number determining the number of components of the wavefunction $\psi$ (having $4^N$ components), written as products of components of Dirac wavefunctions. For $N=1$, Eq. (\ref{LubE}) reduced to the ordinary Dirac equation for the electron, while for $N=2$ the Kemmer equation was recovered in the form given by Belinfante. Lubanski proved that for $N\geq 2$ the representations for $\Gamma_\mu$ became reducible: ``on peut dire alors que l'Eq. [(\ref{LubE})] ne d\'ecrit plus une seule particule mais une superposition de particules'', as for the Kemmer equation. However, ``dans le cas de $N\geq 3$ la situation est ancore plus compliqu\'ee. Les particules d\'ecrites par l'Eq. [(\ref{LubE})] ont non seulement diff\'erents nombres quantiques de spin mais aussi diff\'erentes masses.'' Lubanski,  indeed, recognized also the type of spin and mass spectrum which would be obtained.

\subsection{Rarita-Schwinger equation for spin-3/2 particles (1941)}

\noindent In 1941, W. Rarita and J. Schwinger \cite{rarita} again considered the problem of ``simplifying'' the complicated spinor formalism of the Dirac-Fierz-Pauli for half-integral spin, by treating, in particular, the special case of spin-3/2 particles (already studied explicitly by Fierz and Pauli \cite{fp}). These are described by a spinor field $\psi^\mu$ with an extra vector index, and Rarita and Schwinger succeded to writing down a ``simple'' lagrangian function that ``can be constructed without the intervention of additional fields'', as instead in the Fierz-Pauli case, namely:
\be \label{RSL}
{L} = - \ov{\psi}^\mu \left( \gamma^\tau \partial_\tau + k \right) \psi_\mu - \frac{1}{3} \, \ov{\psi}^\mu \left( \gamma_\mu \partial_\nu + \gamma_\nu \partial_\mu \right) \psi^\nu + \frac{1}{3} \, \ov{\psi}^\mu \gamma_\mu \left( \gamma^\tau \partial_\tau - k \right) \gamma_\nu \psi^\mu .
\ee
From the Euler-Lagrange equation for such a lagrangian, the following equation can be deduced (not reported explicitly in the paper):
\be \label{RSE}
- \left( \gamma^\tau \partial_\tau + k \right) \psi_\mu - \frac{1}{3} \, \left( \gamma_\mu \partial_\nu + \gamma_\nu \partial_\mu \right) \psi^\nu + \frac{1}{3} \,  \gamma_\mu \left( \gamma^\tau \partial_\tau - k \right) \gamma_\nu \psi^\mu = 0.
\ee
Of course, as pointed out by the authors in the general case, additional spurious components are present in the 16-component vector-bispinor wave function $\psi^\mu$, and subsidiary conditions are required even in the ``simpler'' formalism.  

The (short) paper ended with the consideration of the zero-mass limit:
\begin{quote}
In the exceptional case of zero rest mass the wace function admits a gauge transformation,
\[
\psi_\mu^\prime = \psi_\mu + \partial_\mu \varphi, \qquad \gamma^\tau \partial_\tau \varphi =0,
\]
which leaves all physical quantities invariant.

The method here presented for developing the theory of spin $3/2$ thus contains many of the features of both the Proca and the Diraac theory \cite{rarita}.
\end{quote}
Only much later the appearance of the Rarita-Schwinger theory, it was discovered \cite{velo} that a subtle inconsistency was present when the interaction with an external potential is introduced, that is the solutions of the equation propagate at velocities exceeding the speed of light (for arbitrarily weak external fields), thus violating the postulates of special relativity.

\subsection{Bhabha general equation for arbitrary spin (1945)}

\noindent The intricate issue of a relativistic wave equation for arbitrary spin was reconsidered, after Dirac-Fierz-Pauli, by H.J. Bhabha in 1944-5 \cite{bhabha44} \cite{bhabha45} from a general point of view. The simple form of the basic equation was assumed to be the following:
\be \label{bha}
\left( p_k \alpha^k + \chi \right) \psi = 0 ,
\ee
where $p_k= i \partial / \partial x^k$ and $\alpha^k$ are four square matrices whose degrees and commutation rules depend on the spin of the particle considered ($\chi$ is a constant).

The work of Bhabha, however, was not limited just to assume a given form of a general equation, but rather his lucid analysis was aimed to find general conditions to be satisfied when describing a particle of arbitrary spin:
\begin{quote}
In order to develop equations for higher spin values one must find some general principles common to all of them. These are:
\vspace{-0.3truecm}
\begin{item}
\item[A.] It can be deduced from the equations that {\it each} component of the wave function satisfies the second-order wave equation [(\ref{KGE})]. This is physically equivalent to the statement that the particle described by the field has in each case only one value of the rest mass (except for sign).
\item[B.] The particle-field is completely described by an equation of the form [(\ref{bha})] without the help of any further subsidiary conditions \cite{bhabha44}.
\end{item}
\end{quote}
The main problem with spin higher than 1 was that the corresponding theory could not satisfy {\it both} properties, and this was very well illustrated by considering the known Dirac-Fierz-Pauli theory:
\begin{quote}
The DFP equations connect two irreducible spinors, and by a suitable transformation can be split into two sets, one of which still connects the two irreducible spinors together, while the other set only involves one spinor and is in the nature of subsidiary conditions. We shall see below that the first set can be written in the form (\ref{bha}) but without the second set the first set is not equivalent to the DFP equations. The second set, consisting of the subsidiary conditions, is necessary in order that it should be possible to derive a second-order wave equation for each component \cite{bhabha45}.
\end{quote}
The removal of the inconsistencies related to the subsidiary conditions, as we have seen above, required the introduction of additional subsidiary spinors \cite{fp}, and this appeared to Bhabha as a loss of  ``elegance'' of the given mathematical formulation.
\begin{quote}
It would, therefore, appear to be more logical to assume that {\it the fundamental equations of the elementary particles must be first-order equations of the form [(\ref{bha})] and that all properties of the particles must be derivable from these without the use of any further subsidiary conditions} \cite{bhabha45}.
\end{quote}
The price for such a choice was that each component of the wave function did not satisfy the Klein-Gordon equation (\ref{KGE}) and, as a consequence, the particle has states of higher rest mass which ``are an essential feature of the theory and {\it cannot} be eliminated by an artifice any more than the states of negative mass in the usual formulations of the theory.'' This is evidently reminiscent of the Majorana's theory of 1932 \cite{EM1932}. Bhabha then developed a theory where the wave function $\psi$  described particles with different values of the mass, but now ``these must all be considered as different states of the same physical entity, just like the positron and electron, since the equations are irreducible'' \cite{bhabha44}. To make this point even clearer, given the admittedly changed point of view, Bhabha explictly predicted 
\begin{quote}
that should particles of spin 3/2 or 2 exist in nature, they would appear each with two possible values of the rest mass, the lower values of the rest mass being the stable ones in each case [...] The states of different rest mass being merely different states of the same particle, transitions from one mass to another would always be possible under the influence of interaction if sufficient energy were available for the purpose'' \cite{bhabha45}.
\end{quote}
The remaining developments of the theory were just mathematical in nature, mainly regarding the matrices $\alpha^k$ appropriate for a particle that may have spins up to $n/2$, the author giving a sophisticated analysis based on the fact that $\alpha^k$ must transform as a 4-vector under elements of the Lorentz group. Indeed, it was known that the mathematical objects appearing in any well-founded theory of an elementary particle should be tensors or spinors which are irreducible, that is, which cannot be split into two or more parts in a relativistically invariant way, so that, correspondingly, that theory would be based on an irreducible set of equations. The $\alpha$ matrices should, then, generate the nucleus\footnote{This is formed by six infinitesimal transformations; the wave function $\psi$ is connected with the representations
of the group $SO(1,4)$.} of the representation which determined the way the wave function $\psi$ transformed under any given transformation of the Lorentz group. Bhabha showed that the problem of finding all the irreducible representations of the $\alpha$ matrices could be connected with the problem of finding the nuclei of all irreducible representations of the Lorentz group in five dimensions, as already pointed out by Lubanski \cite{luba1942}, the solutions of which were known.

Finally, Bhabha pointed out that, although his theory was a multi-mass and multi-spin one, only one equation applied for each value of $n$, so that the particle described by it displayed only the spin $n/2$ in all circumstances, the mass of the particle depending only on the maximum value of the spin.

The general theory by Bhabha was worth of further consideration in the subsequent years (till the present days), being also discussed in many textbooks of 1950s-1960s \cite{corgel}.

\subsection{Further remarks}

\noindent The relativistic wave equations describing particles with given spin, as discussed above, are the basic ones developed before quantum field theory established itself among physicists as the consistent and complete quantum relativistic theory describing particles. Nevertheless, people continued (until recent times) to consider the issue of relativistic wave equations, but the appearance of the Bhabha equation marked the end of a well-defined phase. Indeed, the different subsequent contributions were (and are) aimed either to further study the features and the consequences of the theories outlined above, or to generalize them to novel theoretical framework (such as supersymmetry, and so on). We do not consider here such contributions, but only mention, in the following, few results obtained in the years just after the fundamental papers by Bhabha and describing quite well what were the mentioned future investigating directions.

The first one was contained in a paper by E. Wild \cite{wild}, where the author showed how particular were the cases for spin 1/2 (electrons) and spin 1 (mesons) in the Bhabha equation (reducible, in those cases, to the electron Dirac equation and to the Kemmer equation), just by studying possible extensions to other types of fields, already considered in the general Bhabha equation. Wild deduced that ``subject to the conditions that (1) the equations must contain no subsidiary conditions, (2) either the total energy must be positive definite, or it must be possible to quantize the equations according to Fermi statistics without using an indefinite metric in Hilbert space, no such extension is possible.'' It is evident, just from this paper, how mathematical overtones were starting to enter the theories considered above. Nevertheless, Wild also introduced an interesting physical interpretation regarding the Bhabha equation: ``the results [obtained by Bhabha] are of considerable interest in that some of the equations so obtained describe `composite' particles which have states of different spin and rest mass, but which, in the state of lowest rest mass, have spin one-half. It is suggested that the proton and neutron might be described by one of these equations.''

Bhabha himself, in  a foundational review of 1949 \cite{bhabha49}``on the postulational basis of the theory of elementary particles'', formulated a generalization of his equation for spin 0 and spin 1, that is, a generalization of the Kemmer equation (\ref{KE}), where the $\alpha$ matrices appearing in Eq. (\ref{bha}) did not satisfy the DKP algebra in (\ref{DKP}), but rather the more general one given by:
\bea  
\alpha^i \alpha^j \alpha^k + \alpha^i \alpha^k \alpha^j + \alpha^j \alpha^i \alpha^k + \alpha^k \alpha^i \alpha^j + \alpha^j \alpha^k \alpha^i + \alpha^k \alpha^j \alpha^i & & \nonumber \\
= 2 g^{ij} \alpha^k + 2 g^{jk} \alpha^i + 2 g^{ki} \alpha^j . & &  \label{bhaalg}
\eea
Bhabha realized that the algebra defined by (\ref{bhaalg}) is ``not finite and one would expect there to be an infinite number of inequivalent irreducible sets of matrices satisfying [(\ref{bhaalg})], all except two of which will not satisfy'' (\ref{DKP}). The corresponding wave equation described a particle of spin 1 having only one value of the mass.

Finally, we end our review with the mathematical paper by V. Bargmann and Wigner of 1948 \cite{BW} on ``group theoretical discussion of relativistic wave equations.'' The mathematical derivation of the wave equation for a particle of spin $s$ and mass $m$ went as following. The wave function $\psi$ of s state comprising $N$ independent, free Dirac particles can be written as a product $\psi = \prod_{\nu=1}^N \psi_\nu (\xi_\nu^i, p_{\nu k})$, where $\xi_\nu^i$ give the four spin functions ($i=1,2,3,4$) and $p_{\nu k}$ the 4-momentum of the $\nu$th particle. Each particle would have its own $\gamma$ matrices and mass $m_\nu$, and the wave equation is just the Dirac equation (\ref{DE2}) $\gamma_\nu^k p_{\nu k} \psi = m_\nu \psi$ (the authors used, as was already usual for theoretical physicists, units such that $\hbar=c=1$). Bargmann and Wigner then proposed that the general relativistic wave equation for the single particle of spin $s$ and mass $m$ be derived from this last equation by setting all $m_\nu$ equal to $m$, all $p_{\nu k}$ equal to a single 4-vector $p_k$ and requiring the wave function $\psi$ (written above as a product) to be symmetric in spin labels $\nu$. The wave equation was then:
\be \label{BWE}
\left( \gamma_\nu^k p_k - m \right) \psi_{1,2,\dot \nu \dots N} = 0
\ee
($N=2s$). Such equations are sometimes called Bargmann-Wigner equations in later developments, and represent the final mathematical systematization of the relativistic wave equation for arbitrary spin, as deduced from general Lorentz group invariance (following the idea of Dirac-Fierz-Pauli).

\section{Majorana unknown contributions} 

\noindent We have seen above that Majorana contributed significantly to the issue of relativistic wave equations, two contributions of which having been published and, thus, made accessible to physicists. The {\it Majorana}-Oppenheimer formulation of photon wave mechanics was, instead, brought to the attention of the scientific community only in recent times \cite{recgian} and, even more recently, other investigations and results by the Italian physicist have been discovered on related issues \cite{holefrench}. With the publication of Majorana's study and research personal notes \cite{volumetti} \cite{quaderni}, a number of outstanding contributions in different areas of physics have come to light, and in the following we will focus on three of them related to the topic considered here. All such contributions were obtained before Majorana's visit to the Heisenberg's Institute in Leipzig, in 1933, although more precise dates are not available \cite{quaderni}.

It is remarkable and amazing how, even in this field,\footnote{For a general review, see Ref.s \cite{adp} \cite{EM2}; see also \cite{funfer}.}  Majorana anticipated several later results among those discussed in the previous section, and his own way of reasoning was different from the particular ones envisaged above. Of course, such unknown contributions did not influence the other protagonists of our story, but the opposite is as well true for the results that came later, so that it is interesting to compare the historical path with the completely independent reasoning of Majorana.

Indirect elements of inspiration, already present in the only two papers published in the issue considered \cite{EM1932} \cite{EM1937}, can then lead to different though similar perspectives. Indeed, it is intriguing to observe - as an example - the indirect inspiration of Majorana's infinite-component theory \cite{EM1932} on Bhabha's theory of 1945 \cite{bhabha45}. Both papers dealt with the problem of a relativistic wave equation for arbitrary spin within a similar formalism (different from the Dirac-Fierz-Pauli one), and both started with first-order equations, in a form similar to the Dirac equation, (\ref{DE}) or (\ref{DE2}). Bhabha then followed Majorana in not requiring the Klein-Gordon equation to be satisfied by the components of the wave function and, like Majorana, Bhabha himself obtained a multi-mass equation, where the rest mass depended on the value of the spin of the particle involved. Although the subsequent developments in Bhabha's paper \cite{bhabha45} testify for a completely independent derivation, the discussion of general {\it forthcoming} results, announced in \cite{bhabha44} and as well announced and {\it then} deduced in \cite{bhabha45}, suggests nevertheless a certain already occurred ``metabolization'' of those results, even before their effective (re-)achievement. And, as a matter of fact, only Majorana considered, for the first time, a multi-mass equation for a wave function not satisfying the relativistic energy-momentum relation (\ref{1}). This is, probably, a beatiful example of latent inspiration, and the following explicit results obtained (and not published) by Majorana may serve also as a guide for a deeper understanding of the issue of relativistic wave equations and its theoretical development.

In the following, we give a sketch of three theoretical contributions reported in the {\it Quaderno n.4} \cite{quad4}, where Majorana elaborated relativistic wave equations for 16-component, 6-component and 5-component spinors, respectively. They are discussed here just in the same order as the material appears in the original documents: as we shall see, this is an important issue, and it is evident that the author obtained the corresponding results just in this order, and {\it not} in the simplistic, more obvious, reverse order aimed at studying particular cases for obtaining the general results published in \cite{EM1932} and discussed above (see below).

\subsection{A 16-component equation for a two-particle system}

\noindent As a starting point, Majorana apparently studied, like de Broglie in 1934, a system formed by two particles each obeying the Dirac equation but, differently from the de Broglie case, he consdered only one particle with a non-vanishing mass $m$, while the other one was supposed massless. The reason for such a choice is not evident,\footnote{Just as a conjecture, this study could be related to Fermi's theory of beta decay and, in such a case, the two fermions should have been an electron and a neutrino. However, it seems that the present notes were written by Majorana well before the elaboration (and publication) of Fermi's theory, in December 1933.} but it is nevertheless evident that the physical idea was very different from that of de Broglie. 

In the original manuscript, Majorana wrote directly the ``Dirac'' equation for the system in an explicit $16\times16$ matrix form, but from this it is possible to reconstruct easily the reasoning behind it as follows. The Dirac equations for the two particles were written in the usual form:
\be \label{ME16c}
\left\{ \ba{l}
\dis 
\left[ \frac{W}{c} + \alphavec \cdot \pbvec + \beta \, m c \right] \psi = 0 , \\ \\
\dis \left[ \frac{W^\prime}{c} + \alphavec^\prime \cdot \pbvec^\prime + \beta^\prime \, m^\prime 
c \right] \psi = 0 ,
\ea \right.
\ee
where a prime referred to the quantities for the second particle, with $m \neq 0$, $m^\prime =0$. As the equation describing the system at hand, however, Majorana considered that obtained by summing side by side the two equations in (\ref{ME16c}):
\be 
\left[ \frac{W}{c} +\frac{W^\prime}{c} + \alphavec \cdot \pbvec + \alphavec^\prime \cdot \pbvec^\prime + \beta \, m c \right] \psi = 0 .
\ee
As for de Broglie, two independent sets of (only) 4 Dirac matrices appear, here built as follows:
\be
\alphavec = \left(  \ba{cccc}
\alphavec^D & 0 & 0 & 0 \\
0 & \alphavec^D & 0 & 0 \\
0 & 0 & \alphavec^D & 0 \\
0 & 0 & 0 & \alphavec^D
\ea \right) , \quad
\beta = \left(  \ba{cccc}
\beta^D & 0 & 0 & 0 \\
0 & \beta^D & 0 & 0 \\
0 & 0 & \beta^D & 0 \\
0 & 0 & 0 & \beta^D
\ea \right) ;
\ee

\

\be \ba{ll}
\dis
\alpha^\prime_x = \left(  \ba{cccc}
0 & 0 & 0 & I_4 \\
0 & 0 & I_4 & 0 \\
0 & I_4 & 0 & 0 \\
I_4 & 0 & 0 & 0
\ea \right) , & \quad
\dis \alpha^\prime_y = \left(  \ba{cccc}
0 & 0 & 0 & - i \, I_4 \\
0 & 0 & i \, I_4 & 0 \\
0 & - i \, I_4 & 0 & 0 \\
i \, I_4 & 0 & 0 & 0
\ea \right) , \\ & \\
\dis \alpha^\prime_z = \left(  \ba{cccc}
0 & 0 & I_4 & 0 \\
0 & 0 & 0 & - I_4 \\
I_4 & 0 & 0 & 0 \\
0 & -I_4 & 0 & 0
\ea \right), & \quad
\dis \beta^\prime = \left(  \ba{cccc}
I_4 & 0 & 0 & 0 \\
0 & I_4 & 0 & 0 \\
0 & 0 & - I_4 & 0 \\
0 & 0 & 0 & - I_4
\ea \right), \ea
\ee
where $\alphavec^D, \beta^D$ are the Dirac $4\times 4$ matrices in (\ref{diracmatrices}), and $I_4$ is the $4\times 4$ identity matrix. Notably, and differently from de Broglie, each set of $16 \times 16$ Dirac matrices does satisfy the Dirac algebra\footnote{It is clear that the matrix $\beta^\prime$ does not appear in the equation above, since $m^\prime =0$, and then it was not used by Majorana. Nevertheless, we have chosen to report also its expression for completeness; it is deduced from the obvious requirement of completing the Dirac algebra satisfied also by this matrix.} in (\ref{diracalgebra}), while each matrix of one set commutes with any of the other set. 

The energy-momentum relation satisfied by the solutions of this Majorana equation is, of course, not the simple one in (\ref{1}), but rather a generalization of it to the actual system; it can be written in an easily recognizable form as
\be \label{1gen}
\frac{W^2}{c^2} = \left[ \sqrt{p^2 + m^2 c^2} \pm \sqrt{p^{\prime 2}} \right]^2
\ee
(each solution of this is, obviously, fourfold).

\subsection{Equation for a 6-component spinor}

\noindent As a subsequent step, Majorana then considered a particle described by a 6-component spinor $\psi$ which satisfies a ``standard'' Dirac equation,
\be \label{ME6c}
\left[ \frac{W}{c} + \alphavec \cdot \pbvec + \beta \, m c \right] \psi = 0 .
\ee
It should be noted that, in the original manuscript, the author considered the Dirac equation in presence of electromagnetic interaction (introduced in the usual way through the 4-potential $A_\mu$ by means of the minimal coupling principle), this denoting not just a merely mathematical interest, but then, when he focussed on the crucial point remarked below, he took the limit of no interaction\footnote{To be more precise, he set to zero the vector potential, but maintained a non-zero scalar potential.} (which is, indeed, irrelevant for that purpose). For simplicity, as for previous discussions, we avoid from the beginning the introduction of electromagnetic interaction.

In the original manuscript, it is apparent that Majorana first wrote down the equations in (\ref{ME6c}) for each of the six spinorial components and {\it then} deduced the explicit form of the four $6 \times 6$ Dirac matrices; they are given as follows:
\be 
\ba{l}
\dis
\alpha_x = \left( \begin{array}{cccccc}
 0 & 1/2 & 0 & 0 & -1/2 & 0 \\
 1/2 & 0 & 0 & 0 & 0 & - 1/2 \\
 0 & 0 & 0 & 0 & 0 & 0 \\
 0 & 0 & 0 & 0 & 0 & 0 \\
 -1/2 & 0 & 0 & 0 & 0 & 1/2 \\
 0 & -1/2 & 0 & 0 & 1/2 & 0
\end{array} \right)
\\ \\
\dis \alpha_y = \left( \begin{array}{cccccc}
 0 & i/2 & 0 & 0 & i/2 & 0 \\
 -i/2 & 0 & 0 & 0 & 0 & i/2 \\
 0 & 0 & 0 & 0 & 0 & 0 \\
 0 & 0 & 0 & 0 & 0 & 0 \\
 -i/2 & 0 & 0 & 0 & 0 & i/2 \\
 0 & -i/2 & 0 & 0 & -i/2 & 0
\end{array} \right)
\\ \\
\dis \alpha_z = \left( \begin{array}{cccccc}
 0 & 0 & - 1/2 & -1/2 & 0 & 0 \\
 0 & 0 & 0 & 0 & 0 & 0 \\
 -1/2 & 0 & 0 & 0 & 0 & 1/2 \\
 -1/2 & 0 & 0 & 0 & 0 & 1/2 \\
 0 & 0 & 0 & 0 & 0 & 0 \\
 0 & 0 & 1/2 & 1/2 & 0 & 0
\end{array} \right)
\\ \\
\dis
\beta = \left( \begin{array}{cccccc}
 1 & 0 & 0 & 0 & 0 & 0 \\
 0 & 0 & 0 & 0 & 0 & 0 \\
 0 & 0 & 0 & 0 & 0 & 0 \\
 0 & 0 & 0 & 0 & 0 & 0 \\
 0 & 0 & 0 & 0 & 0 & 0 \\
 0 & 0 & 0 & 0 & 0 & -1
\end{array} \right)
\ea
\ee
Very interestingly, this set of matrices does {\it not} satisfy the Dirac algebra (\ref{diracalgebra}), but rather they do satisfy the DKP algebra in (\ref{DKP}), as can be tested quite easily. For the sake of completeness, it should be pointed out that, in the original manuscript, there is no hint that Majorana was aware of this fact, thus anticipating the findings of Duffin in 1938 and Kemmer in 1939. Nevertheless, it is intriguing that, instead, he was well aware of the Geheniau decomposition of such algebra or, to be more specific, of the fact that the dimension 6 algebra here considered can be decomposed into a dimension 5 algebra and a trivial one-dimensional one. In order to achieve such a result, Majorana first obtained the energy-momentum relation satisfied by the spinor components, which we write simply as
\be \label{fourfold}
\frac{W^4}{c^4} \left( \frac{W^2}{c^2} - p^2 - m^2 c^2 \right) = 0 .
\ee
This includes, obviously, the usual relativistic relation in (\ref{1}), but an additional (fourfold) solution with 
zero energy emerges, just as in the de Broglie theory of 1934. Then, Majorana considered the non-relativistic approximation of his theory at first order, obtaining:
\bea
& & \psi_1 = 0, \nonumber \\
& & \psi_2 = \frac{p_x - i p_y}{2 mc} \, \psi_6 , \nonumber \\
& & \psi_3 = \psi_4 = - \frac{p_z}{2 m c} \, \psi_6 , \\
& & \psi_5 = - \frac{p_x + i p_y}{2 mc} \, \psi_6 , \nonumber \\
& & \left[ \frac{W}{c} - m c - \frac{p^2}{2 mc} \right] \psi_6 = 0  \nonumber ,
\eea
from which he identified the physical components satisfying the ordinary (non-relativistic) energy-momentum relation, thus avoiding the zero-energy solution. Even more importantly, however, he recognized that one spinor component vanishes in the non-relativistic limit, thus obtaining a {\it physical} decomposition of the dimension 6 algebra, as opposed to the {\it mathematical} decomposition of the DKP algebra.

\subsection{5-component equation}

\noindent Majorana, however, did not content with the non-relativistic decomposition, and further developed the idea of a particle described by a 5-component spinor, along the same lines as for the 6-component theory. The wave equation is just as in (\ref{ME6c}) (the remarks above about electromagnetic interaction apply here as well), but with the four $5 \times 5$ matrices given by:
\be
\ba{l}
\dis
\alpha_x = \left( \begin{array}{cccccc}
 0 & 1/2 & 0 & -1/2 & 0 \\
 1/2 & 0 & 0 & 0 & - 1/2 \\
 0 & 0 & 0 & 0 & 0 \\
 -1/2 & 0 & 0 & 0 & 1/2 \\
 0 & -1/2 & 0 & 1/2 & 0
\end{array} \right) ,
\\ \\
\dis
\alpha_y = \left( \begin{array}{cccccc}
 0 & i/2 & 0 & i/2 & 0 \\
 -i/2 & 0 & 0 & 0 & i/2 \\
 0 & 0 & 0 & 0 & 0 \\
 -i/2 & 0 & 0 & 0 & i/2 \\
 0 & -i/2 & 0 & -i/2 & 0
\end{array} \right) ,
\\ \\
\dis
\alpha_z = \left( \begin{array}{cccccc}
 0 & 0 & - 1/\sqrt{2} & 0 & 0 \\
 0 & 0 & 0 & 0 & 0 \\
 -1/\sqrt{2}& 0 & 0 & 0 & 1/\sqrt{2} \\
 0 & 0 & 0 & 0 & 0 \\
 0 & 0 & 1/\sqrt{2} & 0 & 0
\end{array} \right)  ,
\\ \\
\dis
\beta = \left( \begin{array}{cccccc}
 1 & 0 & 0 & 0 & 0 \\
 0 & 0 & 0 & 0 & 0 \\
 0 & 0 & 0 & 0 & 0 \\
 0 & 0 & 0 & 0 & 0 \\
 0 & 0 & 0 & 0 & -1
\end{array} \right) .
\ea
\ee
Again, such matrices satisfy the DKP algebra (\ref{DKP}), and the energy-momentum relation is similar to that in (\ref{fourfold}), with the obvious replacement of $W^4/c^4$ with $W^3/c^3$ (that is, the zero-energy solution is now threefold).

\subsection{Remarks}

\noindent The three contributions, just discussed, achieved by Majorana clearly point out - once more - his mastery and versatility also about the issue of relativistic wave equations. Here, however, it is probably more interesting to dwell a bit about his reasoning behind calculations. 

The starting point was the description of a system of two particles, each following a Dirac equation. Such a system, however, was considered as a single entity, described by a single equation involving a 16-component spinor. The reasoning is similar to that of de Broglie in 1934 \cite{deb1934}, including the procedure adopted to write down the expressions for the $16 \times 16$ Dirac matrices: in both cases, they were obtained in a quite obvious way as tensor product of two independent sets of ordinary $4 \times 4$ Dirac matrices. The differences in the final results, due to different technical choices, and their interpretation are not very relevant for the present discussion. Instead, it is quite interesting the fact that Majorana turned from his two-particle system described by a 16-component wave function to a one-particle system described by only a 6-component spinor. This is, indeed, a crucial point since apparently it is justified only by assuming that Majorana did realize the possible decomposition later realized by Kemmer \cite{kemmer} and then publicized by Pauli \cite{Pauli1941} (see above). Such a conjecture seems strengthened by the further, explicit decomposition of the 6-component description. 

If this is correct, it is evident that the reasoning by Majorana followed exactly the same basic steps later followed by other people, from de Broglie to Kemmer, thus denoting a uniform development of such ideas. Nevertheless, some distinctive features are present in Majorana's work, since here the transition from 16 to 6 (or 5+1) components meant a transition from the Dirac to the DKP algebra, which is particularly notable if we recall that Majorana obtained it just at the level of the wave equation (which he wrote down directly), and not at the level of the abstract matrix algebra (matrices were deduced by Majorana from his equations). Moreover, quite intriguing is as well the mistery of how Majorana obtained his equations or, in other words, of how he obtained the explicit form of the intervening matrices. In fact, while the 16-component theory seems an obvious generalization of the Dirac's one, with the novel matrices again, and obviously, satisfying the Dirac algebra, this is not at all the case for the 6- and 5-component theory, as testified by the fact that the matrices that he {\it deduced} did not satisfy the known Dirac algebra. In his notes for these topics, Majorana did not mention any commutation or anticommutation relation related to the abstract algebra employed, but his alternative reasoning on this was through the (more directly physical) energy-momentum relations. Now, it is well-known that the requirement of (\ref{1}) led, in the Dirac scheme, to the Dirac algebra in (\ref{diracalgebra}), and this could be invoked to explain the results of the Majorana's 16-component theory, which verify the obvious generalization (\ref{1gen}) of the energy-momentum relation to the system considered. Instead, a similar reasoning could certainly not apply to the Majorana's 6- or 5-component theory, both for the fact that the energy-momentum relation in (\ref{fourfold}) is not at all an obvious generalization of (\ref{1}), given the presence of additional zero-energy solutions, and for the fact that Majorana deduced {\it a posteriori} the energy-momentum relation, just by requiring a non-trivial solution to the homogeneous 6- or 5-component wave equations. It is thus unfortunate that no additional information is available on this crucial point.

Finally, it is remarkable that all such general features are present, though in a latent form, in the Majorana's paper of 1932 \cite{EM1932}, where the infinite-component theory predicted a multi-mass equation (as in the 16-component theory) with only four differently dimensioned matrices, whose form and algebra was not deduced {\it a priori} by imposing the energy-momentum relation  (\ref{1}) (as in the 6- and 5-component theory). In this sense, such different works by the same author are closely interconnected each other, and the three contributions reported above probably served as the starting points for the general theory published in 1932.

\section{Pauli and Fierz about Majorana's infinite-component equation}

\noindent As discussed above, the relevance of the Majorana's paper of 1932 resides not only in the position (and a possible solution) of the problem of a quantum relativistic equation for particles with {\it arbitrary} spin, but also in its curious indirect inspiration on several subsequent works. Now, we will briefly consider the direct influence of that paper on leading scientist or, rather, their careful studies about it, not later resulted in known publications. 

Quite unexpectedly, a key role in the {\it understanding} of the Majorana's paper was played by Pauli (probably informed by Heisenberg), who studied it at least from 1939 to 1947 in two distinct phases, around 1940 and in 1947. This is clearly testified by 9 letters\footnote{That is: 
Pauli to Bhabha, 12 April 1940; 
Pauli to Fierz, 3 July 1940; 
Pauli to Fierz, 17 July 1940; 
Pauli to Fierz, 3 September 1940; 
Pauli to Jauch, 1 Novenber 1940; 
Pauli to Fierz, 12 February 1941; 
Pauli to Fierz, 29 March 1941; 
Fierz to Pauli, 17 March 1947; 
Pauli to Fierz, 30 March 1947.} present in Pauli's correspondence, now published in Ref. \cite{Pauli1993}, from which we quote in the following. The reason for such a strong interest is declared in the first letter of this set, that Pauli wrote to Bhabha on 12 April 1940: 
\begin{quote}
I believe in the existence of much more particles than known until now, particularly on particles with arbitrary values of the spin and of the charge. [...] 

My considerations about the particles with higher spins came to some end now. I think that they exist really, but I can fancy that the complication of the theory comes from the assumptions, that one has to describe a set of particles with a finite number of spin values only. May be the matter becomes simpler, if one introduces a priori an infinite set of spin values (compare Majorana [...]).
\end{quote}
Such an indication (that evidently urged Bhabha to think about what he later published in 1945) clearly points out that Pauli only started to study the Majorana's paper at that time, and further insights in it still had to come. Indeed, a full understanding of the Majorana theory resulted to be not so easy to achieve: ``Der Fall unendlich-reihiger Darstellungen der Lorentzgruppe scheint mir jedenfalls noch nicht genugend untersucht. [...] Lesen Sie doch einmal den Majorana!''\footnote{That is: ``Anyway, it seems to me that the case of the infinite-dimensional representations of the Lorentz group has not been studied sufficiently. [...] Read Majorana!'' (Pauli to Fierz, 3 July 1940)} Or, in the more characteristic Pauli's style, when referring to the Wigner's paper \cite{wigner}: ``Wigner hat die Majoranaschen Arbeit nicht verstanden, wie er Mir zugegeben hat.''\footnote{That is: ``Wigner did not understand the Majorana's equations, as he admitted to me.'' (Pauli to Fierz, 29 March 1941)} Pauli himself, however, had to read the Majorana's paper several times before any definite conclusion.

The first problem he envisaged was that of the solutions of the Majorana equations propagating with a superluminal speed (already noted by Majorana himself): 
\begin{quote}
Das Schlimme an seiner Theorie ist, da{\ss} es bei ihm ebene Wellen gibt, die einer imagin\"aren Ruhmasse entsprechen, d.h. Teilchen, die sich immer mit Uberlichtgeschwindigkeit bewegen gemass $E/c=\pm \sqrt{p^2 - p_0^2}$ ($p_0>0$ beliebig, $|p|>p_0$). [...] 

Die Frage w\"are: gibt es f\"ur unendlich viele Eigenfunktionen [...] auch solche Gleichungssysteme (bzw. Nebenbedingungen), bei denen pathologische Losungen mit $(v^2/c^2) -k^2 <0$ ausgeschlossen sind? Die letztere Forderung sollte da eine ahnliche komplizierende Rolle spielen wie bei uns die Forderung positiver Energie bzw. Ladungsdichte.\footnote{That is: "The bad thing with his theory is that his plane waves correspond to an imaginary rest mass, i.e. particles moving always with a velocity greater than that of light, that is $E/c=\pm \sqrt{p^2 - p_0^2}$ ($p_0>0$ arbitrary, $|p|>p_0$). [...] The question would then be: does it exist an infinite number of eigrnfunction [...] for those equations (or constraints) whose pathologic solutions with $(v^2/c^2) -k^2 <0$ are excluded? This last condition plays a role similar to that of requiring a positive energy or charge density. (Pauli to Fierz, 3 July 1940)}
\end{quote}
It is remarkable that Pauli recognized such a problem (and only this as a real problem: ``Ich sehe deshalb auch gar nichts pathologisches in der Majoranaschen Gleichung [...] Das einzige, was mir bei Majorana pathologisch zu sein scheint, sind die Losungen mit im\"aginarer Masse."\footnote{That is: ``I don't see anything pathologic in the Majorana equation [...]. The only thing which seems to be pathologic in Majorana's theory are the solutions with imaginary mass." (Pauli to Fierz, 17 July 1940)}) as pathologic for the Majorana's theory, but nevertheless he continued to study that theory for a long time. Indeed, in a long letter to Fierz of 3 September 1940, Pauli re-derived (in an alternative way) the conclusions obtained by Majorana, even casting the original Majorana equations into a different, equivalent form and comparing them with the Dirac equation. The interesting conclusion about the ``pathologic solutions'' was that their exclusion should be related directly to the Dirac-Fierz-Pauli equations:  ``Es scheint mir, da{\ss} bei Ausschlu{\ss} von unphysikalischen Zust\"anden mit $p_0^2-p^2 <0$ im wesentlichen unsere Gleichungen herauskommen m\"ussen.''\footnote{That is: ``It seems to me that the exclusion of unphysical states with $p_0^2-p^2 <0$ should come essentially from our equations.'' (Pauli to Fierz, 3 September 1940)} Such a feeling was likely the background for considering the Majorana equations as an interesting mathematical problem, but without interesting physical applications (Pauli to Fierz, 3 September 1940). Nevertheless, again, this did not prevent Pauli to further look inside the Majorana theory, even with the help of Bargmann,\footnote{Bargmann published his final results on the unitary representations of the Lorentz group only years later \cite{barg1947} but, evidently, Pauli was aware of some of his results earlier.} searching for possible alternative representations of the Majorana equations in terms of differential operators. This problem, however, was not so easy to solve, and several failures marked Pauli's research: ``Bis jetzt ist es uns nicht gelungen, die Darstellungen mit $0<J<1$ durch Differential-operatoren zu realisieren. Wir wissen aber auch nicht, was der Punkt $J=1$ des Spektrums gruppentheoretisch bedeuten konnte. [...] Das ist also noch ein offenes Problem.''\footnote{That is: ``Up to now we have not been able to obtain representations with $0 <J <1$ using differential operators. And we do not even know what it might mean the point $J = 1$ of the spectrum in group theory. [...] So this is still an open problem. (Pauli to Fierz, 12 February 1941)} However, Pauli did succeeded to obtain several mathematicl results about the Majorana theory, and his first investigations of 1940-1 concluded with the important result that:
\begin{quote}
Die Majoranaschen Gleichungen mit einem $\zeta$-Raum von $\infty$ vielen Dimensionen geben im $(\zeta,\vec{p})$-Raum zu einer {\it reduziblen} unit\"aren Darstellung der inhomogenen Lorentz-Gruppe Anla{\ss}  (die alle Falle $p_0^2-p^2 >0$, $p_0^2-p^2 =0$ und $p_0^2-p^2 <0$ umfa{\ss}t, aber neben den Diracschen Fallen auch andere).\footnote{That is: ``The Majorana equations with an infinite-dimensional $\zeta$-space correspond, in the $(\zeta,\vec{p})$-space, to a {\it reducible} unitary representation of the inhomogeneous Lorentz group (including all the cases with $p_0^2-p^2 >0$, $p_0^2-p^2 =0$ and $p_0^2-p^2 <0$, besides the Dirac case and other cases.'' (Pauli to Fierz, 29 March 1941)} 
\end{quote}

Even though for a short period of time, Pauli again considered the Majorana equations of 1932 in 1947, probably at the request of Fierz, who, in the meanwhile, went further into the mathematical inspection of them. Now, however, the trouble was with the mass spectrum predicted by Majorana, and re-obtained in a different way by Fierz (see Eq. (\ref{rest})),
\be  \label{rest2}
m_\ell = \frac{M}{\ell+ \frac{1}{2}} \, ,
\ee
by using an equivalent form of the Majorana equations. However, ``die Gleichungen sind ja auf jeden Fall in dieser Form unbrauchbar, weil gerade die gro{\ss}en $\ell$ zu kleinen Massen geh\"oren. Jede Kopplung, z.B. mit Strahlungsfeldern, wird deshalb \"Ubergange nach beliebig hohen $\ell$ zur Folge haben.''\footnote{That is: ``In any case, the equations are really useless in this form because, for very large $\ell$, masses are very small. Any coupling, for example with radiation fields, will result into transitions with any large $\ell$.'' (Fierz to Pauli, 17 March 1947)} The attention then shifted to physics problems. However, Pauli cut short any possible subsequent discussion, by recalling the result already obtained in 1941:
\begin{quote}
Es scheint, da{\ss} der Gesichtspunkt der unit\"ren Darstellung der {\it inhomogenen} Lorentzgruppe der Klassifikation der relativistischem Wellengleichungen kraftefreier Teilchen sehr angemessen ist, indem n\"amlich den {\it irreduziblen} Darstellungen gerade {\it bestimmte Werte der Masse und des Spins} entsprechen. Von diesem Standpunkt aus erscheinen z. B. die Majoranaschen Gleichungen als ganz willk\"urlich, weil reduzibel!\footnote{That is: `` It seems that the point of view of the unitary representations of the {\it inhomogeneous} Lorentz group for the classification of relativistic wave equations of particles in the absence of force is very reasonable, namely that to certain {\it irreducible} representations correspond {\it definite values of the mass and the spin}. From this point of view, for example, the equations of Majorana appear completely arbitrary, because they are reducible!'' (Pauli to Fierz, 30 March 1947)}
\end{quote}
The initial confidence in the existence of ``much more particles than known, with arbitrary values of the spin''
finally changed: the appearance of the lucid analysis of Bhabha in 1945 ended the game.

\section{Conclusions}

\noindent Contrary to a very common belief among physicists, we have seen that the issue of the relativistic quantum mechanical description of particles kept theoreticians busy for a long time. Indeed, although the success of the Dirac equation was ruled even before the experimental observation of positrons, due to the successfull predictions about hydrogen atom with respect to those coming from the Klein-Gordon equation, the search for equations describing particles with higher spin lasted from the early 1930s to mid 1940s, regardless of the effective experimental observations of those particles. 

An important source of inspiration was de Broglie theory of a composite photon (described by a 16-component spinor), not so much for its physical content (a photon is composed of an electron-positron pair or, later, by a neutrino-antineutrino pair), but rather for some mathematical background that later originated, on the one hand, the Proca equation for massive spin-1 particles and, on the other hand, the Kemmer equation for spin-1 plus spin-0 particles (described by a 10- plus 5-component spinor). This line of reasoning led to the final settlement of Bhabha in 1945, who developed a general first-order equation for particles with arbitrary spin and studied the conditions under which such a problem could be solved.

The difficult problem of describing particles with spin higher than 1/2 was previously (1936) considered by Dirac, and later (1939) improved and refined by Fierz and Pauli, who adopted an alternative, more general line of reasoning, involving more advanced mathematics.

While both alternatives are diverse expressions of the same Lorentz group, crucial differences manifest in the two theories when requiring (Dirac-Fierz-Pauli) or not (Bhabha) the {\it a priori} validity of the Klein-Gordon equation for each component of the spinor describing the particle. In the first case, indeed, subsidiary conditions should be added {\it ad hoc} for the theory to be fully consistent, while this does not happen for the second case. But, even more importantly, while the Dirac-Fierz-Pauli formalism is able to describe particles with definite mass and spin, the Bhabha equation is a multi-mass and multi-spin equation.

The issue of a general equation for arbitrary spin was, however, posed by Majorana early in 1932, that is just when only the Klein-Gordon and the Dirac equations were known as relativistic particle equations. His solution was in term of an infinite-component theory based on a multi-mass and multi-spin equation, but this cannot be simply considered just as a precursor (though a generalization, with an infinite number of spinor components) of the later Bhabha equation. Indeed, not only the general ideas underlying the Majorana and the Bhabha equations are the same, but, quite interestingly, the evolution of the specific line of reasoning is itself identical.

As we have shown above by making recourse to unpublished documents, even Majorana developed a 16-component theory describing a system of two Dirac fermions, later developed in a less detailed and more involved way by de Broglie, as well as a 6-component and a 5-component theory for a one-particle system, based on the later discovered Duffin-Kemmer-Petiau algebra and its decomposition. Thus, the difference in the specific line of reasoning of Majorana and Bhabha was only in the period of time required for its evolution: probably few weeks in one case (Majorana), while more than ten years in the other one (Bhabha), with the contribution of several authors. It is very interesting the fact that the same line of reasoning has evolved in an identical manner, regardless of the actors involved: in the history of physics there are not so many similar cases that can be studied. Although Bhabha was introduced (in 1940) by Pauli to the Majorana's paper of 1932, the ``intermediate theories'' discussed above were not included at all in that paper (or in any other published one), and only their latent and very indirect influence may be flowed to Bhabha (if any).

On the other hand, the Majorana's paper of 1932 resulted to be very difficult to fully understand (probably, just for its pregnant meaning and latent physical and mathematical content) even to first-order theoreticians like Wigner and Pauli. In particular, Pauli soon enough (in 1940) recognized that, contrary to naive expectations, the Majorana approach with an infinite rather than finite number of components greatly simplified the matter, even with respect to his own (with Fierz and Dirac) theory. However it occurred about one year of intense study with Fierz to detail his understanding of Majorana's theory, including the recognition that Majorana's equations corresponded to a {\it reducible} unitary representation of the Lorentz group (as in the Kemmer case, for example), as witnessed in Pauli's correspondence. And, quite intriguingly, only in 1947 (that is, after the appearance of Bhabha's, Wigner's and Bargmann's papers) Pauli finally declared his preference in equations with definite values of the mass and the spin. This just testifies for the difficulty of the problem at hand and for the depth of Majorana's reasoning and results.

The saga of the relativistic quantum mechanical description of particles with arbitrary spin came to an end with the final mathematical systematization by Bargmann and Wigner in 1948 along the lines of Dirac-Fierz-Pauli (that is, equations implementing irreducible rather than reduciblke representations of the Lorentz group), although later, different mathematical improvements, generalization, etc. of prevous theories exist that come up today. The end of the story was justified by the recognition that relativistic wave mechanics derived its physical relevance just from its incorporation into quantum {\it field theory}, so that subsequent discussions mainly focused on this last framework. 

Nevertheless, contrary to naive expectations, what reviewed here has an enormous potential interest for present day research as, for example, in condensed matter physics. Indeed, without considering the standard case of simple superconductors, where Cooper pairs are described by a scalar field \cite{GL} that, in the Ginzburg-Landau theory, just follows the Klein-Gordon equation (\ref{KGE}), in several materials the charge carriers behave exactly as Dirac fermons \cite{diracfermions}, and a number of key phenomena are just predicted by the Dirac equation (\ref{DE}) applied to these particles \cite{kt}. Moreover, several other investigations suggest that exotic quasiparticle excitations in a variety of interesting condensed matter systems follow, instead, the Majorana equation (\ref{MEST}), that is they are fermionic excitations that are their own antiparticles (Majorana fermions) \cite{majoranafermions}. Other exotic phenomena (such as, for example, that considered in \cite{sr2ruo4})  exist that, in principle, could be described by other equations (such as, for example, the Kemmer equation (\ref{KE}) for describing electrons in these exotic materials grouped in $s$-wave or $p$-wave pairs). 

The subject is, then, worth to be further exploited in current science, with possible novel interesting results to come: by paraphrasing Pauli, it ``has not been studied sufficiently.'' Yet.

\end{document}